\documentclass[numberedappendix]{./emulateapj}
\usepackage{amssymb,amsmath,amsthm}
\usepackage{longtable}
\usepackage{amssymb}
\usepackage{amsmath, xspace}
\usepackage{graphics,graphicx} 
\usepackage{rotating}
\renewcommand{\thefootnote}{\arabic{footnote}}
\renewcommand{\thefootnote}{\arabic{footnote}}

\def\xmm{{\textit{XMM-Newton\xspace}}}
\def\esas{{\textit{XMM-ESAS\xspace}}}
\def\sas{{\textit{SAS\xspace}}}

\begin{document}


\title{High Resolution XMM-Newton Spectroscopy of the  Cooling Flow Cluster Abell 3112}


\author{
G.~Esra~Bulbul\altaffilmark{1},
Randall~K.~Smith\altaffilmark{1},
Adam Foster\altaffilmark{1},
Jean~Cottam\altaffilmark{2},
Michael~Loewenstein\altaffilmark{2},
Richard~Mushotzky\altaffilmark{2}, and
Richard~Shafer\altaffilmark{2}
}
\affil{$^1$ Harvard-Smithsonian Center for Astrophysics, 60 Garden Street, Cambridge, MA~02138.\\
$^2$ NASA Goddard Space Flight Center, Greenbelt, MD, USA.\\}
\email{ebulbul@cfa.harvard.edu}

\begin{abstract}

We examine high signal to noise \xmm European Photon Imaging
Camera (EPIC) and Reflection Grating Spectrometer (RGS) observations to determine the physical characteristics of the gas
in the cool core and outskirts of the nearby rich cluster A3112.  The XMM-Newton Extended Source
Analysis Software data reduction and background modeling methods were used to analyze the \xmm EPIC data. From the EPIC data we find that the iron and silicon abundance gradients show significant increase towards the center of the cluster while the oxygen abundance profile is centrally peaked but has a shallower distribution than that of iron. The
X-ray mass modeling is based on the temperature and deprojected density distributions of the intra-cluster medium determined from EPIC observations. The total mass of A3112 obeys the M $-$ T scaling relations found using \xmm and \textit{Chandra} observations of massive clusters at $r_{500}$. The gas mass fraction $f_{gas}= 0.149^{+0.036}_{-0.032}$ at $r_{500}$, is consistent with the seven-year WMAP results. The comparisons of  line fluxes and flux limits on the Fe XVII and Fe XVIII lines obtained from high resolution {\it RGS} spectra 
indicate that  there is no spectral evidence for cooler gas associated with the cluster with temperature below 1.0 keV in the central $<$ 38$''$ ($\sim$ 52 kpc) region of A3112.  High resolution {\it RGS} spectra also yield an upper limit to the turbulent motions in compact core of A3112  (206 km s$^{-1}$). We find that the energy contribution of turbulence to total energy is less than 6 per cent. This upper limit is consistent with the amount of energy contribution measured in  recent high resolution simulations of relaxed galaxy clusters. 
\end{abstract}
%%%%%%%%%%%%%%%%%%%%%%%%%%%%%%%%%%%%%%%%%%%%%%%%%%%%%%%%%%%%%%%%%%%%%%%%%%%%%%%%%%%%%%%%%%%%%%%%%%%
\section{Introduction}

Galaxy clusters represent the largest scales of organized matter in
the Universe. The intra-cluster medium (ICM) traces the cluster
gravitational potential  created primarily by dark matter. It emits
X-rays due to the highly ionized gas which has been heated by infall
from the intergalactic medium and enriched by early supernovae.
Although hydrostatic models produce temperature and density profiles
in good agreement with observations, in the most centrally condensed
cases the core emission rates are high
enough to radiate a substantial fraction of the gas thermal energy
over the cluster lifetime \citep{fabian1994}. Metal ions
({\it e.g.} iron, silicon, and oxygen) recombine as the temperature
falls, resulting in large increases in the specific
cooling rate and producing an effective run-away to a nearly neutral
gas below $10^4$\,K and are associated with burst of star formation \citep{mcdonald2011}. 
 Assuming pressure equilibrium, a continual
infall of gas into the core accompanies the temperature drop, which
over the cluster lifetime, would be responsible for a large fraction of
the cD galaxy mass. This general picture, dubbed as a ``cooling
flow", was also inferred in smaller organized structures (groups) and
even large elliptical galaxies.  Observations with the
\textit{Einstein Observatory} showed almost a third of all X-ray
clusters showed significant apparent cooling flows \citep{white1997}.

Understanding what happened to the cool gas in cooling flow clusters
was one of the prime problems in extragalactic astrophysics until
analysis of the \xmm\ Reflection
Grating Spectrometer (RGS)  spectra showed that they do not, in fact,
cool down to the runaway temperature
\citep{peterson2003}.  This result shifted attention
to more difficult question: ``What terminates the cooling
flow?'' A number of models have been proposed; see \cite{peterson2006}
for a general review of understanding of cluster core observations and
models.  
\begin{table*}[ht!]
\centering
\caption{\it   \xmm Observations of A3112}
\scriptsize
\begin{tabular}{lcccccccc}
\hline \hline 
 	& Instrument & Obs Id  & R.A. & Decl & Exposure Time  &  Clean Time&Obs Date\\ 
 	&&& (J2000) & (J2000) &(ks)&(ks)&&\\
\hline
\\
\xmm &RGS, EPIC& 0603050101	& 03 17 57.4& - 44 14 14.7& 118.9 &102.2 & 2009 July 21	\\ 
\xmm & RGS, EPIC& 0603050201	& 03 17 57.4& - 44 14 12.8& 80.8 & 80.5 & 2009 Aug 08	\\
\\ 
\hline
\label{table:previousObs}
\vspace{5mm}
\end{tabular} 
\end{table*}

From short early observations of putative cooling
flow clusters, \citet{peterson2003} inferred that there was a
`floor' to the temperature of the cooling gas at $\sim 1/3$ the temperature of the gas, $T_{\rm
max}$, just outside the `cooling radius'
where the cooling time was on the order of 0.2 or less of the Hubble
time.  The analysis of {\it RGS} spectra found
less than 10 per cent of the cooling rates expected from the observed X-ray surface brightness
profiles. Recent analyses of high-quality spectra of cooling flows
\citep{sanders2011,takahashi2009,sanders2008,werner2006a,werner2006b,peterson2003} have shown that the distribution of emission
measure versus temperature does not have a universal form as expected in constant pressure cooling flows and that
there is indeed gas in some systems over a much wider temperature
range. The wide temperature range in the Centaurus cluster shows that
there exists gas with a cooling time of less than $10^7$\,years \citep{sanders2008}. On
the other hand in HCG 62 there is a very narrow range of
temperatures \citep{sanders2010b}. The physical reason for these differences is as yet
unknown and may provide the missing piece to the cooling flow
puzzle. The abundances of the elements in some of these high signal-to-noise
spectra are also unusual, for example in Centaurus the ratio of oxygen (O), magnesium (Mg) and nickel (Ni) to iron (Fe) (with respect to solar value) grows monotonically,
from  O/Fe$\sim $1/3 solar to Ni/Fe$\sim$2.5 solar value.
These extreme non-solar abundance patterns may hold clues to the
nature of star formation in these systems \citep{odea2008} 
and the origin of the energy that prevents run-away cooling.

\begin{figure}[h!]
\centering
\vspace{5mm}
\includegraphics[width=8cm]{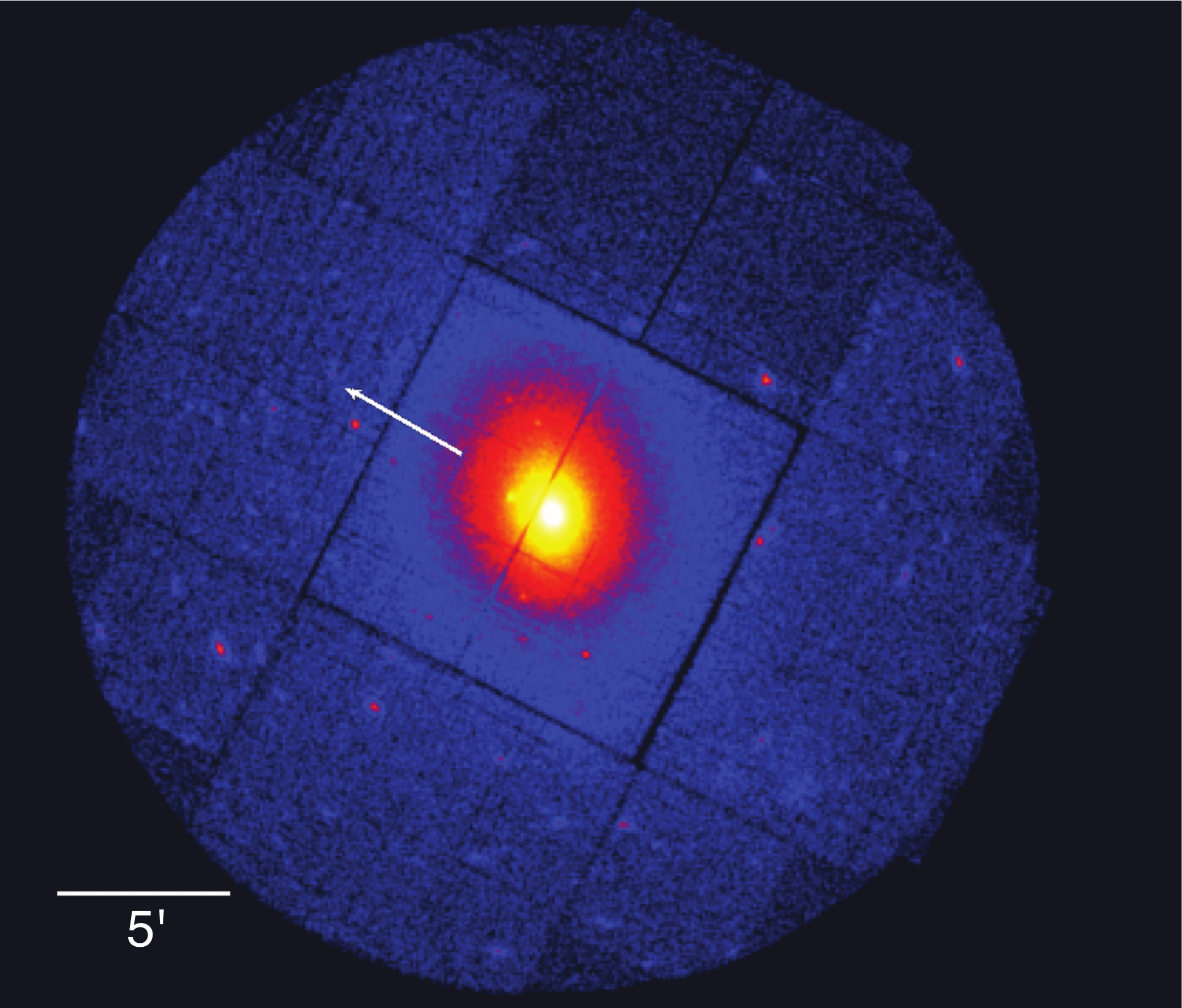}
\vspace{5mm}
\caption{XMM-Newton image of A3112. The white arrow shows the dispersion direction for {\it RGS} observations. The 5$'$ scale corresponds to 410.25 kpc at the cluster's redshift.}
\label{fig:MOS1image}
\end{figure}

A3112, at a redshift of 0.075, is a rich X-ray cluster,  with
a powerful radio source, PKS 0316-444, in the center \citep{takizawa2003}. 
ASCA \citep{edge1992} and ROSAT \citep{allen1997} observations showed that A3112 has a strong apparent cooling flow, with an inferred mass deposition rate of 
$\dot{M}\ \sim$ 415 M$_{\odot}$yr$^{-1}$. As implied by ASCA results, the bright central cD 
galaxy ESO 248-G006 is interacting with the surrounding ICM \citep{takizawa2003}. 
They also detected disturbances in the ICM around two radio ``lobelike'' regions near the center of A3112 (r $\sim$ 10$''$) from Chandra observations, indicating that the X-ray gas and radio lobes are interacting. Finally, \citet{takizawa2003} also found a temperature decrease and abundance increase toward the center of the cluster and placed an upper limit to the contribution of X-ray emission from a cooling flow of 10\% of the total emission.

In addition to these results, a soft X-ray excess emission has been reported from \xmm  and Chandra and Suzaku observations  of A3112 \citep{nevalainen2003,bonamente2007,lehto2010}. \citet{bonamente2007} argue that the excess X-ray emission is of a non-thermal nature and can be explained with the presence of a population of relativistic electrons with $\sim$7\% of the cluster's gas pressure. They also noted that this soft excess emission is equally well fit by a non-thermal power-law model or by a thermal model of $\leq$0.62 keV gas temperature and zero metal abundance. 
\citet{lehto2010} showed that the excess emission is $\sim$10\% of the cluster hot gas emission at 0.4$-$1.7 keV
assuming that the emission above photon energies of 2 keV originates purely from the ICM based on Suzaku and XMM-Newton observations.

In this paper we examine the results of the deep (182 ksec) \xmm European Photon Imaging Camera (EPIC) and the RGS observations  of  A3112. 
This paper is organized as follows. \S2 describes the \xmm EPIC and {\it RGS} data processing and background modeling methods. In \S\ref{sec:epicResults}, we determine Fe, Si, S, and O  abundance profiles and masses of A3112 out to $r_{500}$  based on the EPIC spectra products. In \S\ref{sec:rgsResults}, we present the results of  {\it RGS} observations. We place an upper limit to the  turbulent gas motions and the mass deposition rate as a function of temperature. We discuss the implications of our results and provide our conclusions in \S\ref{sec:conclusion}. 
Errors correspond to the 90\% confidence level throughout the paper unless otherwise noted.
In the mass calculations,  we assume the cosmological parameters h = 0.73, $\Omega_{M}$= 0.27 and $\Omega_{\Lambda}$ = 0.73.

\begin{figure*}
\centering
\includegraphics[angle=-90,width=14cm]{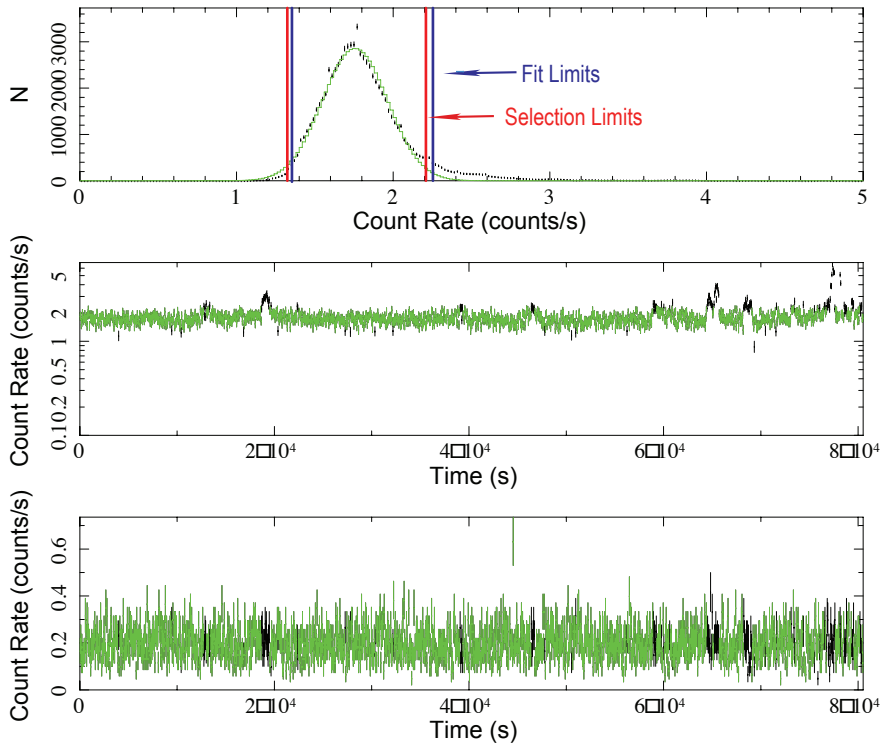}
\caption{Light curve and light curve histograms of {\it MOS1} obtained from observation 0603050201.
The top panel shows a Gaussian shaped curve fit to the histogram where there is flaring characteristics
due to soft proton contamination.The blue vertical lines indicate the range for the Gaussian fit, while the red vertical lines show the upper and lower bounds for filtering the data. The green curve shows the Gaussian fit to the data. The middle and bottom panels show the intervals excluded due to soft proton contamination (in black) in the FOV and corner light curves.}
\vspace{5mm}
\label{fig:lightCurve}
\end{figure*}

\section{Observations and Data Analysis}

\subsection{XMM EPIC Observations}
\label{sec:xmmEpicDataAnalysis}

A3112 was observed with \xmm for 119 ks in 2009 July  and 80 ks 
in 2009 August (see Table \ref{table:previousObs}). The EPIC 
data processing and background modeling were carried out with the \xmm Extended Source Analysis Software (\textit{XMM-ESAS}) and methods \citep{kuntz2008,snowden2008}. 
We processed the MOS data with the \xmm Science  Analysis System (SAS) version 10.0.0 and  {\it PN} data  with version 11.0.0. The basic filtering and calibration were applied to event files using  \esas tools \textit{epchain}, \textit{emchain}, \textit{mos-filter}, and \textit{pn-filter}.
The tasks \textit{pn-filter} and \textit{mos-filter} call the \sas task \textit{espfilt} to filter the data for obvious soft proton flares (SP), using the filtered events files created by \textit{emchain} and \textit{epchain}. The task \textit{espfilt} creates two light curves for the MOS: one in the field of view
(FOV) and one from the unexposed corners of the instrument in a high energy band from 2.5 keV to 12 keV.
Then the high-energy count rate histogram, which should have a  roughly Gaussian profile (possibly with a high count rate tail) was created from the FOV light curve. The task fits
a Gaussian to the peak of the distribution and removes flares 
which exceed a threshold count rate set at $\pm1.5\sigma$ level, using the best-fit Gaussian parameters. This was used to create good-time-interval file. 
In Figure \ref{fig:lightCurve}, we show the time periods removed  (in black) in our analysis due to SP flaring. The exposure time after filtering is shown in Table \ref{table:previousObs}.

\begin{figure}[h!]
\centering
\includegraphics[angle=-90,width=8.5cm]{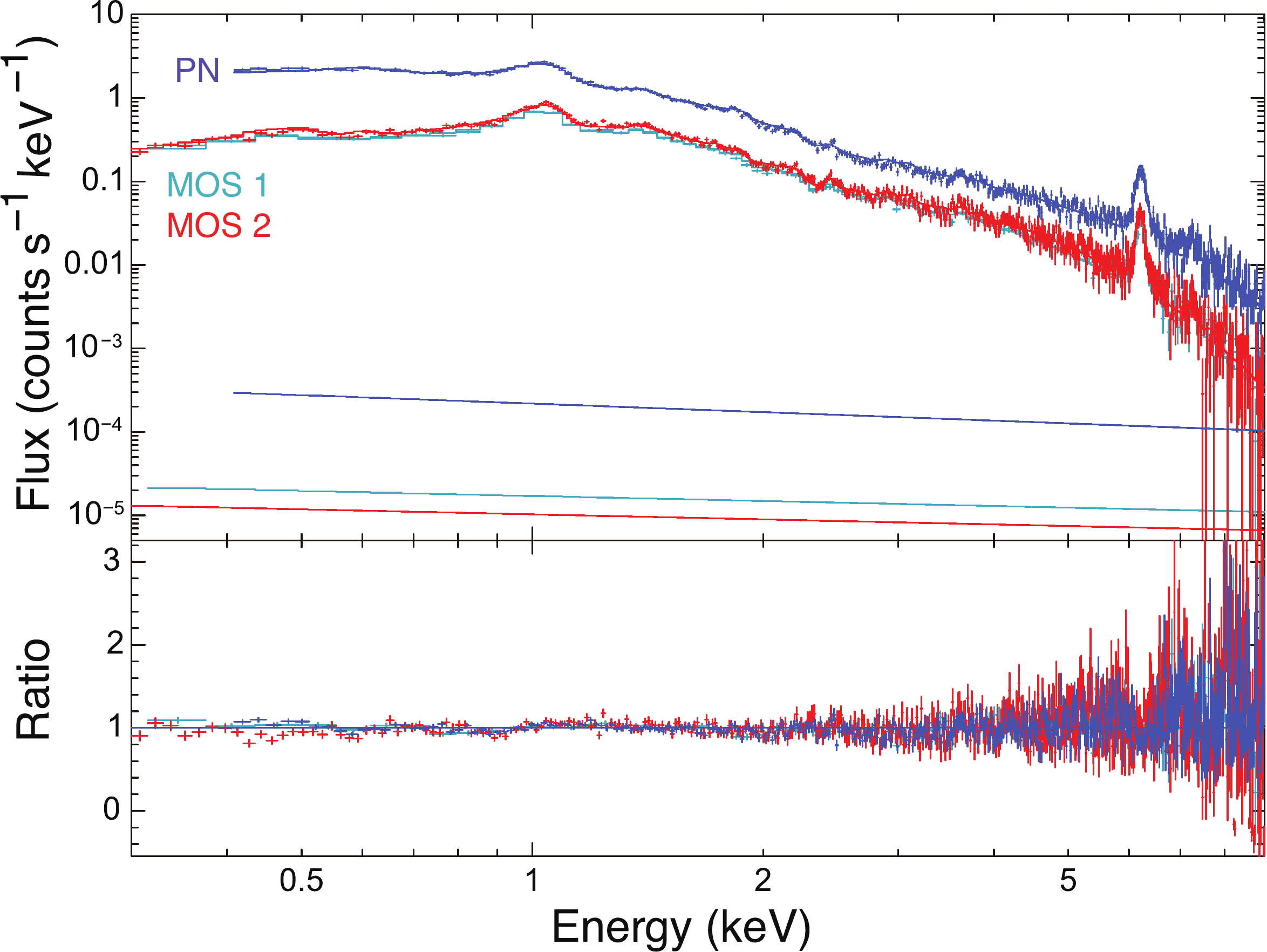}
\caption{An example fit {\it MOS1} (turquoise), {\it MOS2} (red), and {\it PN} (blue) spectra  with residual plot from the innermost 30$''$ region of A3112 obtained from the observation 0603050201. The diagonal lines  represent the contamination due to residual SP background. The fit was performed as described in \S\ref{sec:xmmEpicDataAnalysis} with the model background subtracted. The best fit temperature and abundance 
are 3.30 $\pm$ 0.02 keV and 1.08 $\pm$ 0.08 respectively.}
\label{fig:Spectrum-0-30as}
\vspace{5mm}
\end{figure}

Spectra were extracted in a series of contiguous, concentric annuli centered at the X-ray centroid after the point sources were removed using the automated point 
source detection \esas task \textit{cheese-bands}. The products were examined visually to correct for the central point sources that were missed by the task \textit{cheese-bands} due to the separation limit we used in the task. The center of the image was determined using \textit{xmmselect} in detector coordinates for each observation.
We used the \esas tasks \textit{mos-spectra} and \textit{mos-back} to create the spectra in the 0.3$-$10.0 keV energy band and model the particle background and instrument response for the MOS detectors. 
\textit{pn-spectra} and \textit{pn-back} were used to create {\it PN} spectra and background files in the 0.4$-$10.0 keV energy band. The resulting MOS and {\it PN} spectra of the innermost 30$''$ region are shown in Figure \ref{fig:Spectrum-0-30as}.

Proper treatment of the background, along with cross talk effects, are the keys to XMM-Newton observations of extended sources. The proper methods used in our XMM-Newton MOS and {\it PN} data analysis are described below.

\subsubsection{Instrumental Background} 
 
The instrumental background consists of two components: instrumental fluorescent lines and SP background.
The instrumental fluorescent lines Al K$\alpha$ (E $\sim$ 1.49 keV) and Si K$\alpha$ (E $\sim$ 1.75 keV)  in 
the MOS and the Al K$\alpha$ line  (E $\sim$1.49 keV) from the {\it PN} spectra were modeled 
with two Gaussian components.

Although the light curve filtering  was applied during reduction process as described in \S\ref{sec:xmmEpicDataAnalysis}, some SP residuals may still be present  in the data.
To minimize the effect of SP residuals we model the residual SP contamination 
with a single  power law model using diagonal spectral redistribution matrices
obtained from \esas Current Calibration Files database \footnote{$http://xmm.vilspa.esa.es/external/xmm\_sw\_cal/calib/ccf.shtml$}. The power-law index was left free to 
vary between 0.2 and 1.3 with a free normalization factor for the innermost spectra. The normalizations of the power-law model of the other annuli were tied to the normalization of power-law of the innermost region according to their area. The power-law background models for {\it MOS1} (black), {\it MOS2} (red), and {\it PN} (green) data are also shown in Figure \ref{fig:Spectrum-0-30as} for the innermost 30$''$ region.

 \subsubsection{Cosmic X-Ray Background }
 
The cosmic X-ray background (CXB) is spatially  variable  over the sky and has multiple components. 
The CXB was modeled by two thermal components for the emission from the Local Hot Bubble (LHB) and the Galactic halo, and a non-thermal component for the unresolved cosmic sources.
We use the ROSAT All$-$Sky Survey (RASS) spectrum to model the CXB using the background tool at the High Energy Science Archive Research (HEASARC) Web site. The RASS spectrum was extracted from an annulus surrounding the cluster center between 1$^{\circ}$ and 2$^{\circ}$, with the assumption that the annulus spectrum reasonably represents the CXB  in the direction of the cluster.
We simultaneously modeled the emission from the LHB (or heliosphere) with a cool 
unabsorbed single temperature thermal component (E $\sim$ 0.1 keV) and the Galactic hotter halo and intergalactic medium with an absorbed thermal component (E $\sim$ 0.2 keV).
 
Another CXB component arises from the emission from unresolved point sources. 
The contamination due to unresolved cosmological sources is modeled using an absorbed power law
component with a power law index, $\alpha$ $\sim$1.46 \citep{snowden2008}.

  \begin{table}[ht!]
\caption{\it O VII and OVIII emission surface brightness}
\centering
%\normalsize 
\begin{tabular}{cccc}
\hline \hline 
Instrument & Line		 & Energy 		& Line Surface Brightness	\\
		&			& (keV)		& (LU)		\\
\hline\\
MOS			&  OVII		& 0.56	& 3.92 $\pm$1.05\\
MOS			& OVIII		& 0.65	&1.75 $\pm$ 0.56\\
MOS+PN		& OVII		& 0.56	&1.77 $^{+0.61}_{-0.72}$ 		\\
MOS+PN		& OVIII		& 0.65	&1.66 $^{+0.38}_{-0.61}$  	\\
\\
\hline
\label{table:oxygenFluxes}
\end{tabular} 
\end{table}

\subsubsection{Solar Wind Charge Exchange}  

The cosmic soft X-ray background below  $\sim$1 keV is dominated by line emission
from oxygen ions, such as the 0.56 keV O VII (the O VII triplet from n = 2 $\rightarrow$ 1) and 0.65 keV OVIII (the O VIII Ly $\alpha$ transition). The origin of these lines could be the hot gas in the LHB, charge exchange between Oxygen ions in the solar wind (SWCX) and geocoronal or interplanetary material, or a combination of the these two processes. \citep[e.g.,][]{snowden2004, smith2007,fujimoto2007}.
The potential SWCX emission was modeled by two Gaussian components for lines of OVII at E $\sim$ 0.56 keV and OVIII at E $\sim$ 0.65 keV with widths set to zero.
To distinguish between the LHB X-ray emission and the SWCX emission, we calculate line fluxes 
in units of photons cm$^{-2}$ s$^{-1}$ sr$^{-1}$ (hereafter LU, for ``line units''; see Table \ref{table:oxygenFluxes}).
While relatively large, these values are not outside the range seen in other observations; see, e.g., \citet{koutroumpa2011}. Without high-resolution 
non-dispersive spectroscopy determining  the true origin of this emission is  not possible.

\begin{figure}
\centering
\includegraphics[angle=-90,width=9.5cm]{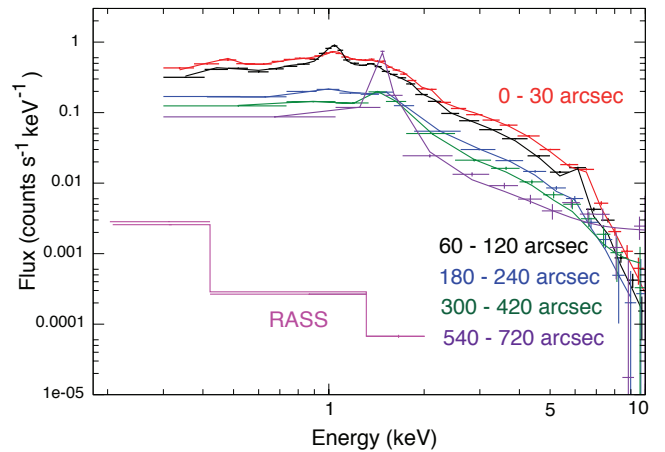}
\caption{60 {\it MOS1}, {\it MOS2} and {\it PN} spectra obtained from observations listed in Table \ref{table:previousObs} were fit simultaneously using a method described in \S \ref{sec:xmmEpicDataAnalysis} with a RASS spectrum. The total reduced $\chi^{2}$ obtained from this fit is 1.28 for 14650 degrees of freedom.  We plot five {\it MOS1} and a RASS spectra surrounding the A3112 centeroid obtained from observation 0603050201 as an example.}
\label{fig:EPICSpectra}
\vspace{5mm}
\end{figure}

\subsubsection{Cross Talk}

Due to the finite point spread function (PSF) of the \xmm EPIC detectors, significant numbers of photons originating from one region may be detected in another region if the region annuli are comparable to the size of the PSF. This is an especially important source of bias in temperature measurements
of strong cool core clusters, such as A3112 \citep{snowden2008}. The annuli widths of 0.5$'$ and 1$'$ we used in this work require the use of a  crosstalk correction as an additional model component in the spectral analysis. The cross-talk effect leads the measured temperature of the innermost annulus to be higher and of the outer annulus to be lower than the true value. We used the \sas task \textit{arfgen} to create cross-talk ancillary region file (arf) to correct for photons originating in one annulus and detected in another annulus.

 \begin{table}
\caption{\it Elemental Abundances Obtained from EPIC Observations. Upper limits are 90 per cent.} 
\vspace{-2mm}
\begin{center}
\begin{tabular}{lcccc}
\hline \hline 
Region 	&  Fe	& Si	& O 	& S \\
(arcsec)	& (Solar)	& (Solar)& (Solar)	&(Solar)	\\
\hline\\
0 - 30	& 1.11$\pm$0.01	& 0.94$\pm$0.07	&0.82$\pm$0.04 &0.87$\pm$0.09\\
30 - 60	& 0.54$\pm$0.02	& 0.12$\pm$0.11	&0.53$\pm$0.08&$<$ 0.17\\
60 - 120	& 0.41$\pm$0.02	& $<$ 0.11		&0.37$\pm$0.12 & $-$ \\
120 - 180	& 0.32$\pm$0.03	& $-$			&0.36$\pm$0.15 & $-$\\
180 - 240 & 0.34$\pm$0.04	& $-$			&0.57$\pm$0.22 & $-$\\
240 - 300 & 0.35$\pm$0.05	& $-$			&0.36$\pm$0.23 & $-$\\
300 - 420 & 0.24$\pm$0.04	& $-$			&0.19$\pm$0.21 & $-$\\
\\
\hline
\label{table:}
\end{tabular} 
\end{center}
\end{table}

\begin{figure*}[ht!]
\centering
\includegraphics[angle=-90,width=13cm]{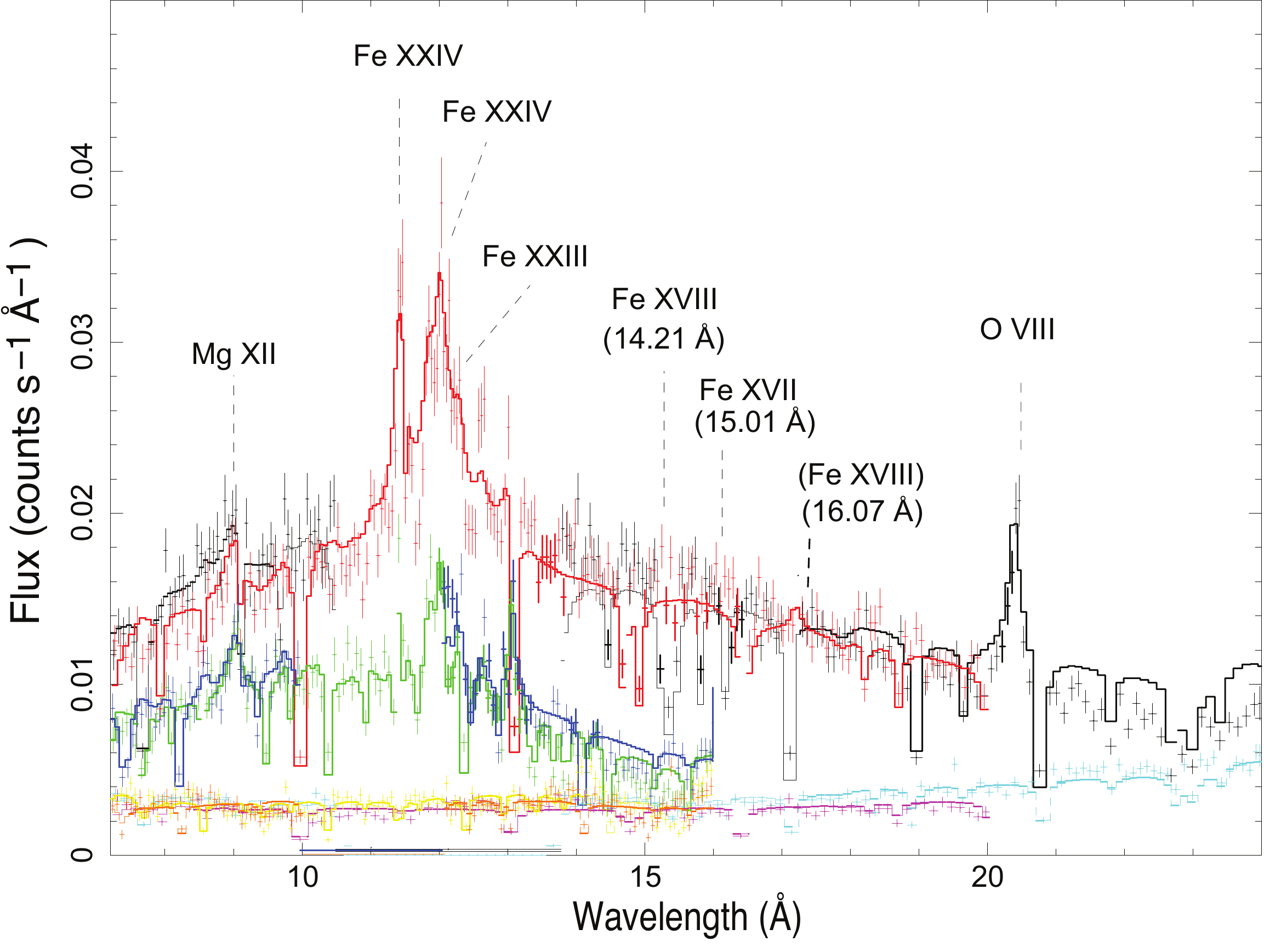}
\caption{ 1st order and 2nd order {\it RGS} spectra and background fit with a single temperature thermal model. The best fit temperature and abundances are 3.82 $\pm$ 0.21 keV  and  0.78 $\pm$ 0.02 respectively. The  $\chi^{2}$ of the source spectra is 4149.5 for 2732 degrees-of-freedom and background fit is 2167.1 for 1332 degrees-of-freedom.} 
\label{fig:rgsSpectra}
\vspace{5mm}
\end{figure*}

\subsubsection{EPIC Spectra Fitting}

The fitting was done in the spectral fitting package XSPEC 12.6.0 \citep{arnaud1996}. The 0.3$-$10 keV  energy band was used for {\it MOS} fits whereas 0.4$-$10 keV band was used for {\it PN} fits. The Galactic absorption column density was left free. The spectral fitting was done simultaneously for a total of 61 spectra using the extended C$-$statistics. 
The diffuse cluster X-ray emission was fit with an absorbed single-temperature collisional  equilibrium plasma model ({\it apec}) \citep{smith2001} and AtomDB v2.0.1 (A. Foster
et al. 2011, in preparation). The final $\chi^{2}$ value was obtained from the best fit determined by C-statistics. Since C-statistic does not provide a direct estimate of the goodness-of-fit and computing the null hypothesis probability distribution of simultaneously fit 61 spectra requires a large amount of computational power over a long period of time we assessed the goodness of the fit using the $\chi^{2}$ value we calculated from the final fit obtained from C-statistics.
We found that total reduced was $\chi^{2}$ of 1.28 for 14650 degrees of freedom (dof) from this simultaneous fit. The temperature and abundance parameters were constrained to be the same for {\it MOS1}, {\it MOS2} and {\it PN} detectors but allowed to vary for each region.  Abundances were given with respect to the solar values in \citet{anders1989}. The normalizations of the single temperature thermal model for the cluster emission were linked for the {\it MOS} and {\it PN} observations. The temperature for the CXB model was linked for different  regions and  instruments. The OVII and OVIII line parameters representing the SWCX were linked for {\it MOS} and {\it PN} data. Figure \ref{fig:EPICSpectra} shows the five {\it MOS1} spectra and a RASS spectrum obtained from the final fit as an example.

\subsection{{\it RGS} Observations }
\label{sec:RGSProcessing}

The high resolution grating spectrometer (RGS) provides high-resolution X-ray spectroscopy in the wavelength range from 5 to 38 \AA.  
We examined two {\it RGS} observations (0603050101 and 0603050201, see Table \ref{table:previousObs}) of A3112 using  the version 10.0 of the Science Analysis Subsystem ({\it SAS} ) software. The {\it SAS}  task \textit{rgsproc} was used to create the filtered event file and source spectrum. Time periods where the background count-rate exceeds a threshold count-rate value 0.2 counts s$^{-1}$ were removed.
The light curves of background events created from the filtered event files showed no sign of SP flaring.  The exposure times after the filtering process are shown in Table \ref{table:previousObs}.

\begin{figure*}[ht!]
\centering
\includegraphics[angle=-90,width=8.6cm]{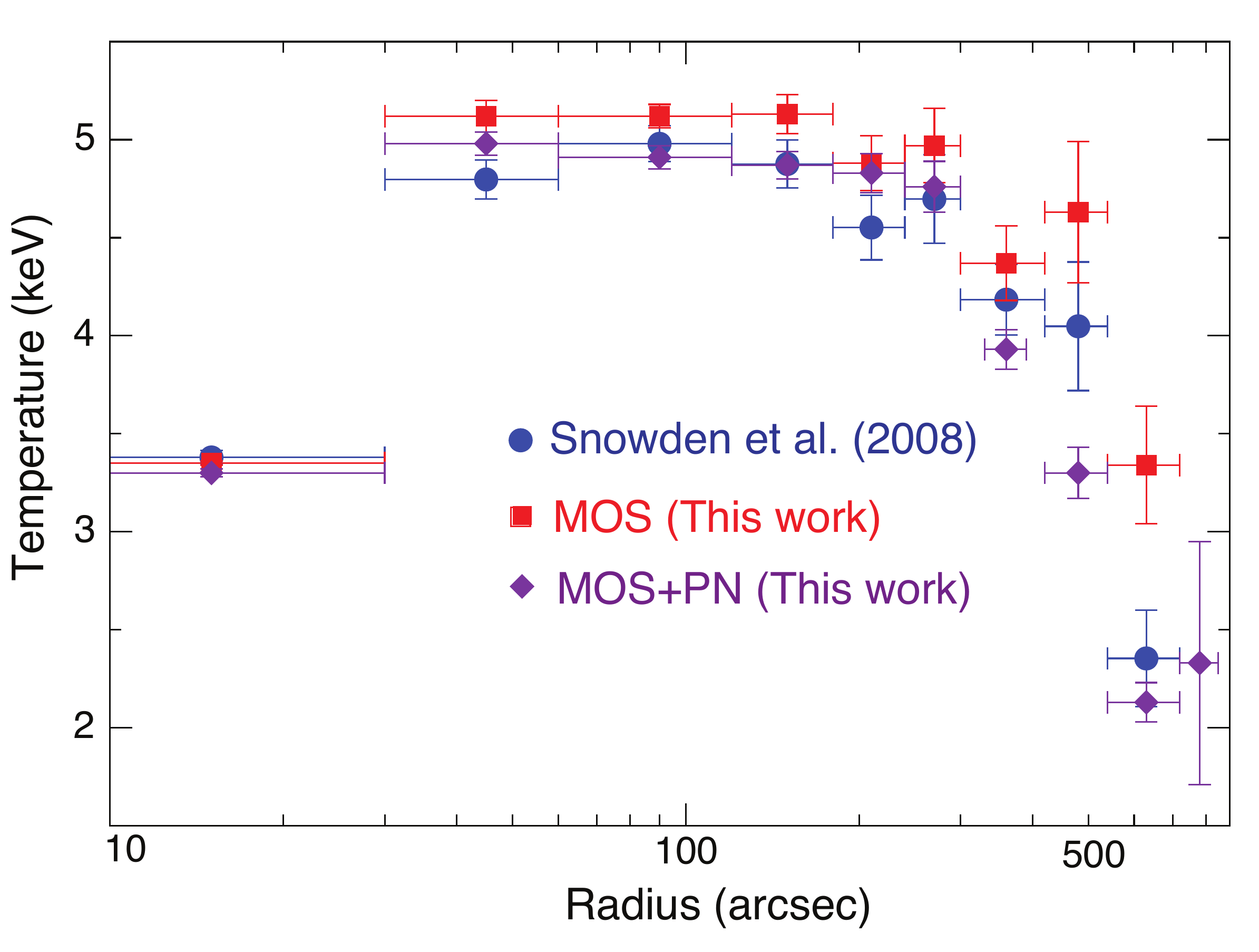}
\includegraphics[angle=-90,width=8.6cm]{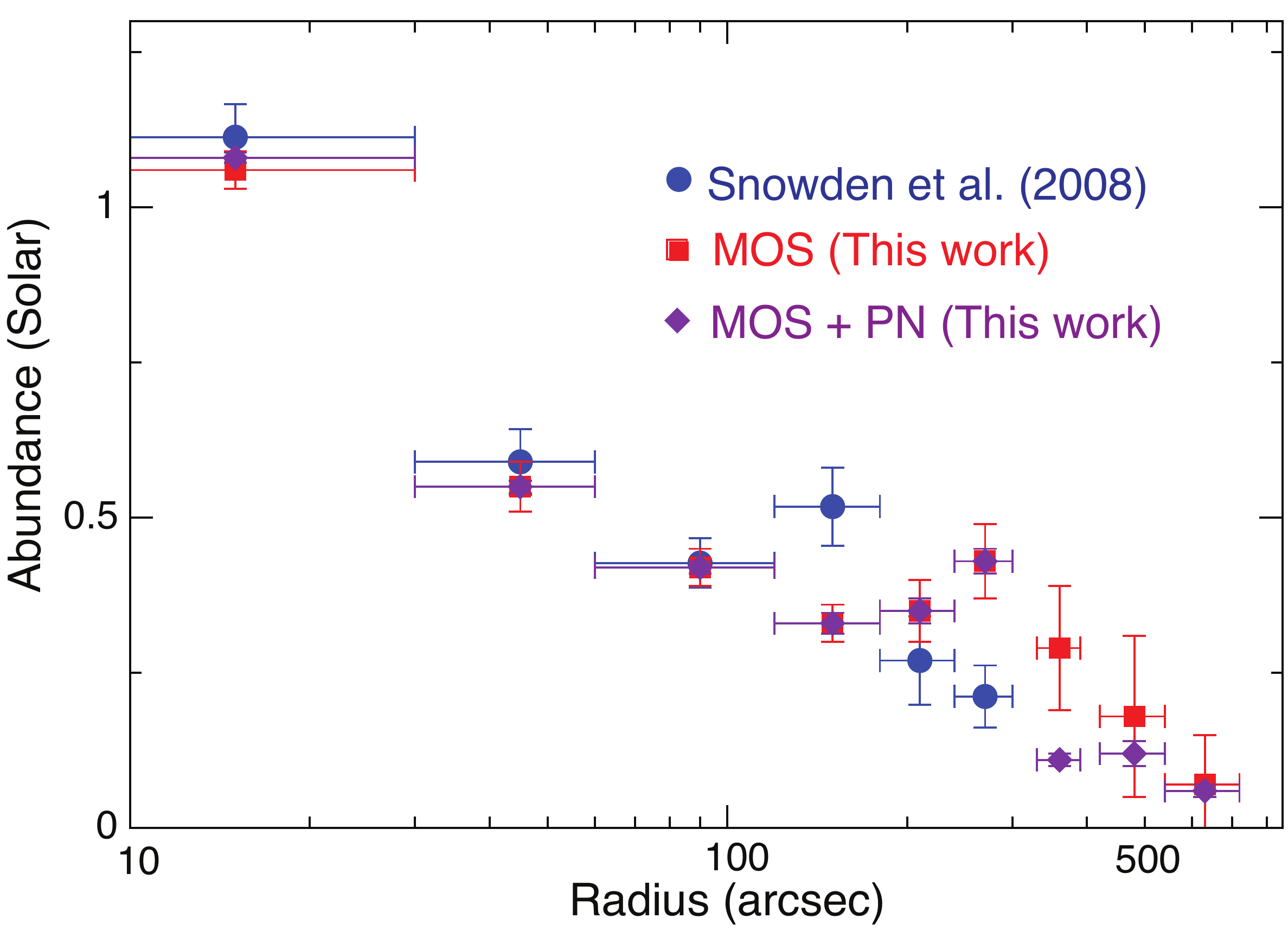}
\caption{Radial profiles of temperature and abundance for A3112 obtained from the single temperature thermal model (\textit{apec}) fit. The temperature and abundance measurements  were obtained  from the {\it MOS1}, {\it MOS2}, and {\it PN} data are shown in purple.  The {\it MOS1} and {\it MOS2} results are indicated in red. The blue data points are the measurements reported in \citet{snowden2008}. }
\label{fig:epicTAP}
\vspace{3mm}
\end{figure*}

 {\it RGS} spectra were extracted by making selections both in the spatial (dispersion/cross-dispersion) and energy (dispersion/PI) planes. The default  90 per cent of the PSF of a point source selection region  in the cross-dispersion direction was used for the extraction of source spectra including 90 per cent of the peak of the pulse-height distribution. We generated the {\it RGS} background and response matrix files using the {\it SAS}  tool \textit{rgsbkgmodel} and  \textit{rgsrmfgen}. The RGS1 spectra obtained from two observations (0603050101 and 06030501201) listed in Table \ref{table:previousObs} were co-added using the task \textit{rgscombine}. The same process was applied for RGS2 observations. The combined {\it RGS} 1st order and 2nd order spectra were then fit in the 7 $-$ 28\AA\ and 7 $-$ 16\AA\ wavelength range respectively (see Figure \ref{fig:rgsSpectra}).

Due to the slitless nature of {\it RGS} spectrometers the observed line emission from an extended source is broadened according to the spatial extent of the source. The observed broadening in wavelength ($\Delta \lambda$) is

\begin{equation}
\Delta \lambda =  \frac{0.138}{m} \Delta \theta \AA
\label{eqn:broadEqn}
\end{equation}

\noindent where  $\Delta\theta$ is the source extent in arc-minutes and \textit{m} is the diffraction order. \footnote{$http://xmm.esac.esa.int/external/xmm\_user\_support/\linebreak documentation/uhb/rgsspecres.html$} The slitless {\it RGS} spectrometers do not allow one to distinguish a change in X-ray photon wavelength and a change in position along the dispersion direction. The most straight-forward way to correct for the line smearing due to source extend is to use {\it SAS} routine \textit{rgsrmfsmooth} prior to spectral fitting. 
The routine modifies the response matrix by convolving it with a spectral model based on the angular structure function computed from a high-resolution {\it MOS} or {\it Chandra} images in the {\it RGS} dispersion direction.
This approach is based on a simple approximation that the X-ray spectrum is uniform and the spatial distributions of all the lines are identical \citep{rasmussen2001}.

\section{EPIC  Results}
\label{sec:epicResults}

The temperature and abundance profiles, shown in Figure \ref{fig:epicTAP}, obtained by fitting 61 {\it MOS} and {\it PN} spectra are obtained from both observations (see Table \ref{table:previousObs}) with a single temperature thermal model (\textit{vapec}). The oxygen (O), silicon (Si) and iron (Fe) and sulfur (S) metallicities were allowed to be free in the fit with the other elemental abundances tied to Fe, except He which was fixed to solar.

\begin{figure}
\centering
\includegraphics[angle=0,width=9.5cm]{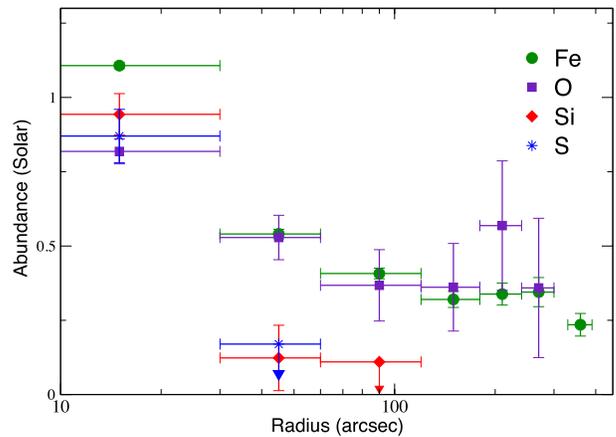}
\caption{Radial distributions of Iron (Fe), Silicon (Si), Sulfur (S) and Oxygen (O) abundances with 90 per cent confidence limits obtained from the XMM-Newton {\it MOS} and {\it PN} observations. The best fit values are obtained by fitting {\it MOS} and {\it PN} spectra with an absorbed \textit{vapec} model. 
}
\label{fig:epicFeOSi}
\end{figure}

We compare our temperature and abundance measurements
with those reported by \citet{snowden2008} (red data points in Figure \ref{fig:epicTAP}). 
Our temperature and abundance profiles obtained from combined {\it MOS} and {\it PN} fits are consistent with those measured  by \citet{snowden2008}. The differences in the outskirts is mainly due to the exposure time difference between this work (200 ksec) and \citet{snowden2008} (20 ksec). {\it MOS} temperature measurements are systematically higher than {\it MOS}+{\it PN} combined temperatures shown in the left panel of Figure \ref{fig:epicTAP}.
A systematic difference of up to 7\% in the wide band temperatures of other clusters between {\it MOS}  and {\it MOS}$+${\it PN} detectors was noted by \citet{nevalainen2010}. The abundance measurements obtained from {\it MOS} observations are in good agreement with those obtained from {\it MOS}+{\it PN} observations and the \citet{snowden2008} results.

Figure \ref{fig:epicFeOSi} shows the radial distribution of Fe, Si, S and O elemental abundances obtained from the {\it MOS} and {\it PN} spectra. The fit was performed with an absorbed single temperature thermal model (\textit{vapec}). The Fe, Si and O elemental abundances were allowed to vary independently while carbon (C), sulfur (S) and argon (Ar) were coupled with Fe abundance  as we were unable to obtain good constraints for the abundance of these elements.  We find that the Fe abundance shows a significant increase towards the center of the cluster while the O abundance profile is centrally peaked but has a shallower distribution than that of Fe.
Si was detected only in innermost 60$''$.  Nevertheless, there is also an indication of an increase in the Si abundance in the cluster core. Type Ia supernovae (SNe Ia) produce significant amounts of Fe  and Si while type II supernovae (SNe II) produce large quantities of O but very little Fe and Ni. The centrally peaked trend of Fe is  qualitatively consistent with the idea that the contribution from SNe Ia relative to that from SNe II increases towards the cluster center \citep{kaastra2001,tamura2004}. Fe is still being added to the ICM specifically in the core by the SNe Ia, and O is a well-mixed product of the SNe II  that explode soon after the first star bursts. The centrally peaked distribution of O may be due to ram-pressure stripping of gas from in-falling galaxies \citep{dePlaa2007}. The comparison of metal abundances  obtained from EPIC and {\it RGS} observations with current SNe models will be addressed in the upcoming paper (G. E. Bulbul \& R. Smith (2012, in preparation).

\begin{figure*}[ht!]
\centering
\includegraphics[angle=-90,width=8.5cm]{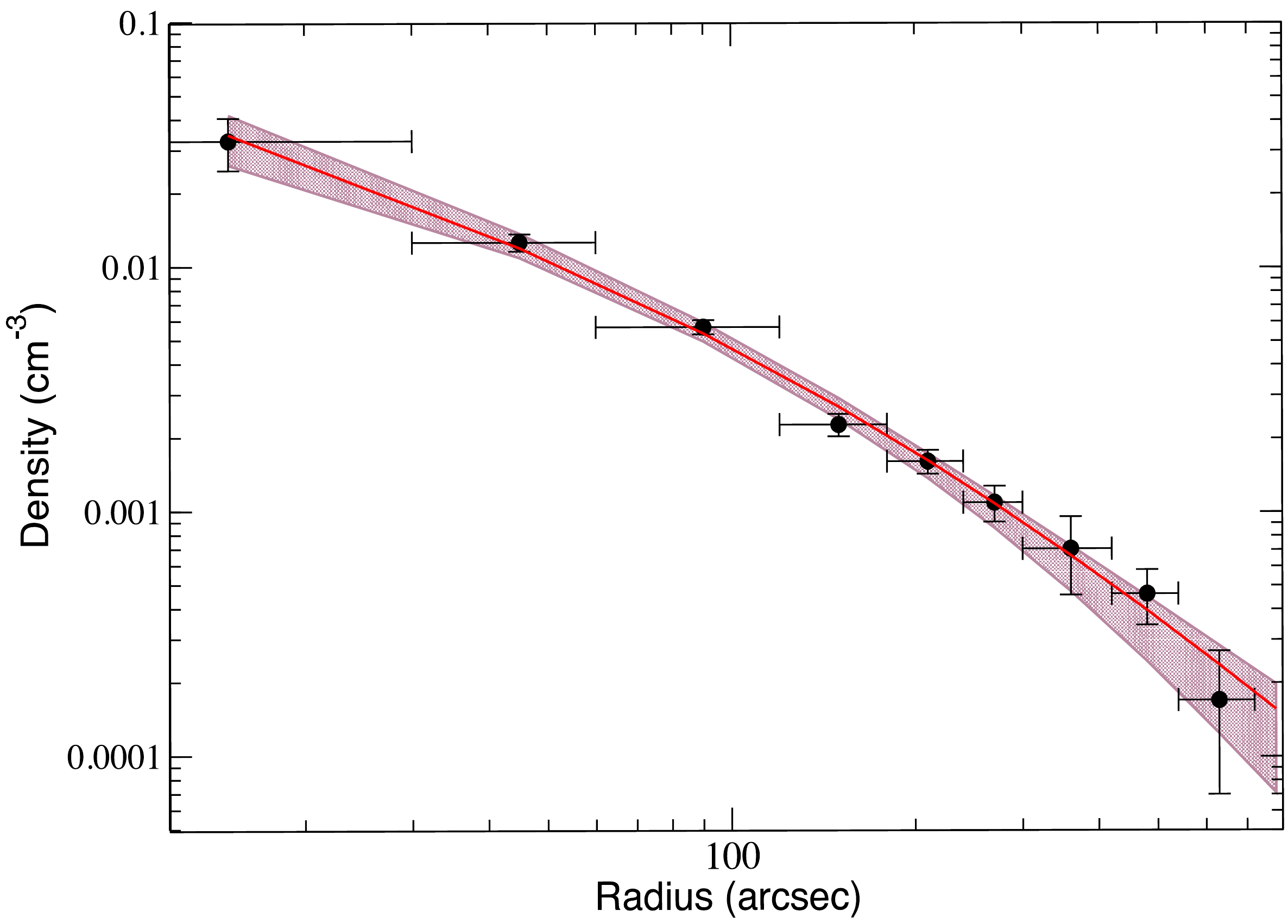}
\hspace{4mm}
\includegraphics[angle=-90.0,width=8.0cm]{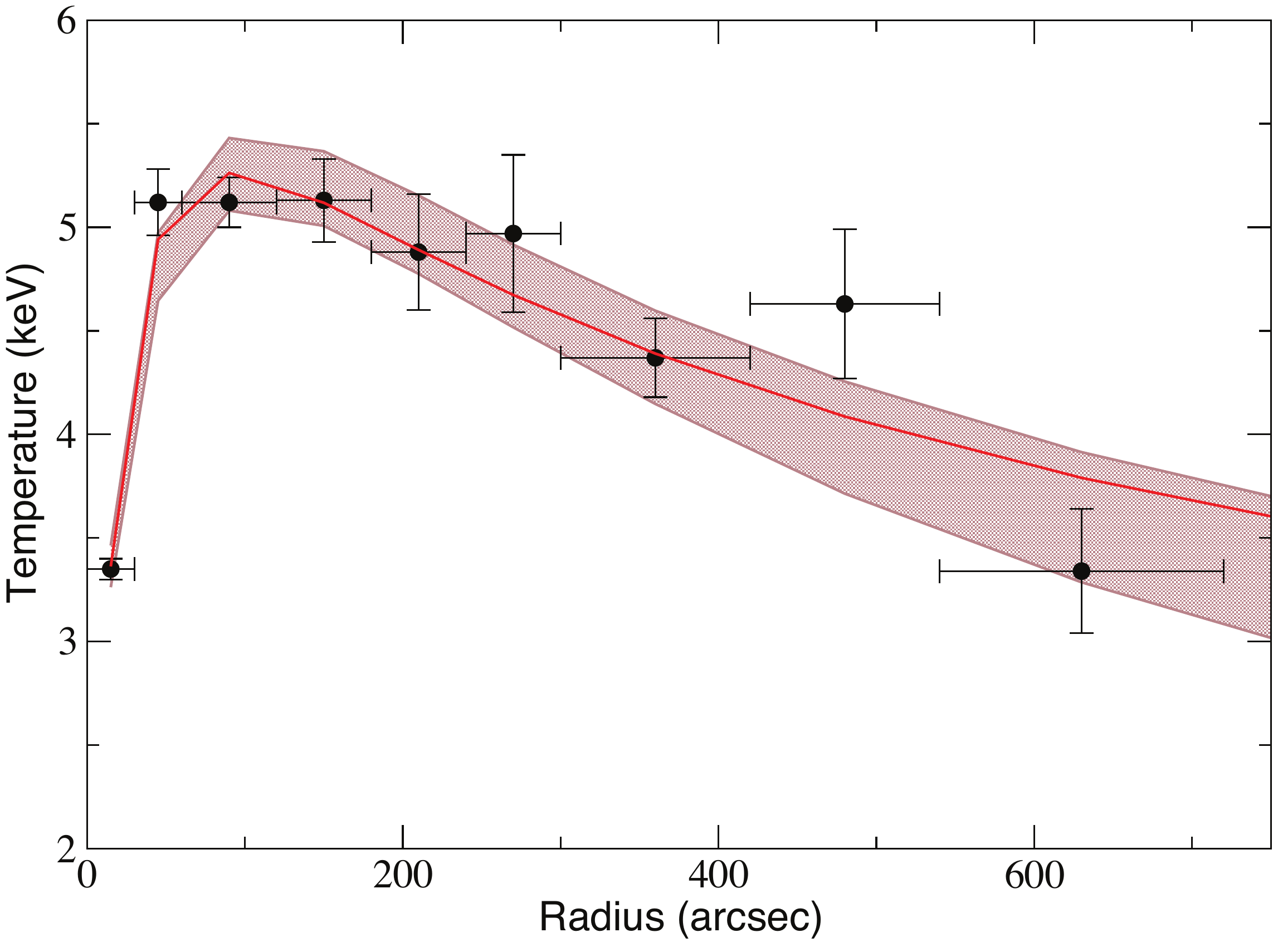}
\caption{Deprojected electron number density and temperature profiles for A3112. Left  panel: the number density profile where the black points are derived from the normalization of the single temperature thermal fit from XMM-Newton {\it MOS} spectra, the red line and the shaded area shows the best fit model and the 90 per cent confidence limit obtained from the model fit. Right panel:  the black data points are the temperature measurements obtained from {\it MOS} spectra. The red line shows the best fit spectroscopic-like temperature profile obtained from the projection of \citet{bulbul2010} model and the shaded region shows the 90\% confidence level obtained from the model fit. The combined $\chi^{2}$ of the fit is 13.3 for 12 degrees of freedom.}
\label{fig:fit}
\end{figure*}

 \subsection{Radial Profiles of Cluster Masses}
 
In order to examine the cluster properties as a function of radius, we obtained the projected density from the normalizations of the best-fitting thermal plasma model to {\it MOS} data we used in \textit{XSPEC}. The normalization is  related to the emission measure as

\begin{equation}
Norm = \frac{10^{-14}}{4\pi(D_{A}(1+z))^{2}}\int n_{e}n_{H}dV,
\label{eqn:xspecNorm}
\end{equation}

\noindent where $n_{e} \sim1.2\ n_{H}$ for a fully-ionized cosmic plasma and $D_{A}$ is the angular diameter distance.
Instead of correcting for projection by combining a series of models, we use  a model-independent approach to subtract the projected spectra from each successive annulus to produce a set of deprojected spectra  \citep{fabian1981, kriss1983}. This  direct `onion peeling' deprojection method  assumes a spherical geometry and that  the density is constant within each annulus. The projected normalizations of the single temperature thermal model (\textit{APEC}) were deprojected using the `onion peeling'  method. These  deprojected normalizations were then used to generate the deprojected number density profile using Equation \ref{eqn:xspecNorm}. 

We describe the density distribution of the hot plasma in galaxy clusters using the model proposed by \citet{bulbul2010}.  The deprojected density profile was represented by

\begin{equation}
\begin{aligned}
n_{e}(r)=n_{e0} \left(\frac{1}{(\beta-2)}\frac{(1+r/r_{s})
^{\beta-2}-1}{r/r_{s}(1+r/r_{s})^{\beta-2}} \right)^{n}\tau_{cool}^{-1}(r)\\
\end{aligned}
\label{eqn_coolcore_gas_density}
\end{equation}

\noindent where  $n_{e0}$ is the normalization factor for the density profile and \textit{n} is the polytropic index.
The core taper function $\tau_{cool}$, used for cool core clusters such as A3112 is

\begin{equation}
\tau_{cool}(r)=\frac{\alpha+(r/r_{cool})^{\gamma}}{1+(r/r_{cool})^{\gamma}}
\end{equation}

\noindent where $r_{cool}$ is the cooling radius, $\alpha$ and $\gamma$ determine the strength 
and shape of the cooling in the cluster center.

\begin{table*}
\caption{\it Best-fit Parameters of the \citet{bulbul2010} Model} 
\begin{center}
\begin{tabular}{cccccccccccc}
\hline \hline 
 $n_{e0}$	&  $\beta$ 	& $n$ 	& $r_{s}$ & $r_{s}$& $r_{cool}$ & $r_{cool}$&$\alpha$ & $\gamma$& $T_{0}$	\\
 ($10^{-2}$ cm$^{-3}$)&	&		& (arcsec)	& (kpc)& (arcsec)	& (kpc)	& 	& & (keV)		\\
\hline\\			
1.17$^{+0.66}_{-0.32}$&1.83$^{+0.29}_{-0.27}$& 3.60$^{+1.68}_{-0.78}$& 64.9$^{+48.8}_{-26.2}$& 88.7$^{+66.7}_{-35.8}$ &84.5$^{+35.8}_{-35.5}$ & 115.5$^{+48.9}_{-48.5}$  &0.03& 0.73$^{ +0.15}_{-0.07}$&  14.9$^{+4.9}_{-5.3}$ \\
\\
\hline
\label{table:bestFitB10}
\end{tabular} 
\end{center}
\end{table*}
\citet{bulbul2010} model also provides the temperature profile, which is 
 
 \begin{equation}
T(r)=T_{0}\left(\frac{1}{(\beta-2)}\frac{(1+r/r_{s})^{\beta-2}-1}{r/r_{s}(1+r/r_{s})^{\beta-2}} \right)\tau_{cool}(r).
\label{eqn:coolCore_tempProf}
\end{equation}

\noindent where $T_{0}$ is the normalization factor for the temperature, $r_{s}$ and ($\beta-1$) are the scaling radius and the slope of the total matter distribution.

In order to fit the temperature profile obtained from {\it MOS} observations, we projected the three-dimensional temperature model shown in Equation \ref{eqn:coolCore_tempProf} along the line-of-sight. This projection approach is based on a weighting method that correctly predicts a good approximation of the spectroscopic temperature for a mixture of different temperature and abundance plasma \citep{mazzotta2004,vikhlinin2006}. The spectroscopic-like 
temperature $T_{sl}$ is,

\begin{equation}
T_{sl}=\frac{\int _{V}w\ T^{-1/2}\ d\it{V}}{\int_{V} w\ T^{-3/2} \ dV}
\label{eqn:Tsl}
\end{equation}

\noindent where $w$ is the weighting factor 
$w= n_{e}^{2}\ T^{-\alpha}$  and $\alpha$ is $\sim$ 0.75 for  the \xmm and Chandra  observations  \citep{mazzotta2004}. The spectroscopic-like temperature obtained using the three dimensional temperature profile (Equation \ref{eqn:coolCore_tempProf}) was then compared to the temperature profile obtained from {\it MOS} observations. The result of this fit is shown in left panel of Figure \ref{fig:fit}.

The fitting was performed using the Monte Carlo Markov Chain algorithm, with Metropolis-Hastings sampling, to determine posterior distributions for 
the best fit model parameters. The best-fit parameters and 90 per cent confidence intervals of the model are shown in Table \ref{table:bestFitB10}.
We obtained a $\chi^{2}=13.3$ for 12 dof for the combined temperature and density profile fits. 
 
The gas mass $M_{gas}$ is computed by integrating the gas density profile within the volume,

\begin{equation}
M_{gas}(r) =  4\pi \mu_{e} \ m_{p}\int n_{e}(r) \ r^{2} \ dr
\end{equation}

\noindent where $\mu_{e}$ and $m_{p}$ are the mean molecular weight and the proton mass. The total mass is 

\begin{equation}
\begin{aligned}
M(r)=\frac{4\pi\rho_{i}r_{s}^{3}}{(\beta-2)}\left( \frac{1}{\beta-1} +\frac{1/(1-\beta) - r/r_s}{(1+r/r_{s})^{\beta-1}}\right)\tau_{cool}(r).
\end{aligned}
\label{eqn:gen_nfw_totalMass2}
\end{equation}

\noindent where $\mu$ is the mean molecular weight and $m_{p}$ is the proton mass. The normalization factor for the total matter density is $\rho_{i} = (T_{0} k (n + 1)(\beta- 1))/(4\pi G \mu m_{p}rs^{2})$. The gas mass fraction is 

\begin{equation}
f_{gas} = \frac{M_{gas}}{M_{tot}}.
\end{equation}

\begin{figure}
\centering
\includegraphics[angle=-90,width=8.5cm]{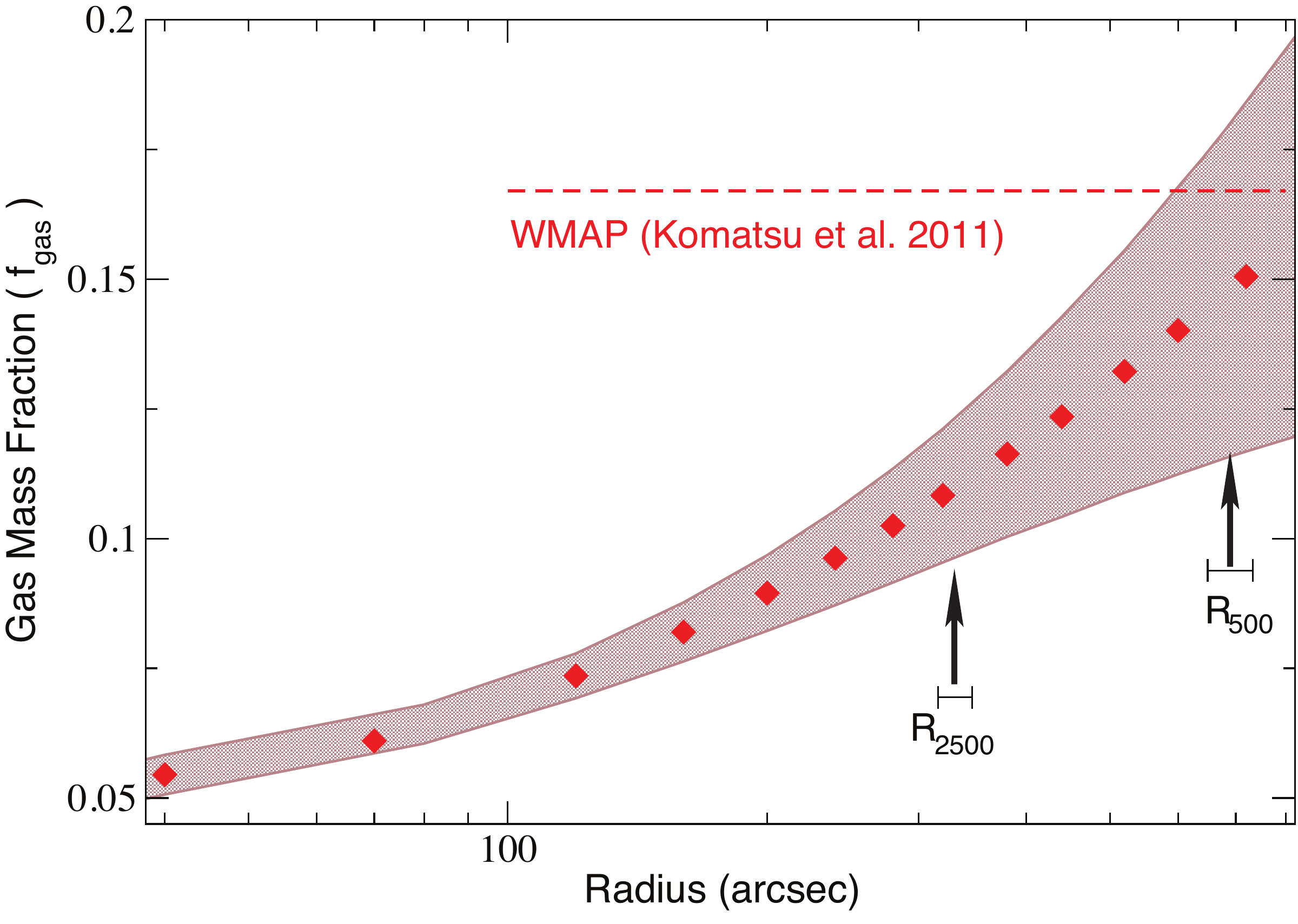}
\caption{Gas mass fraction profile of A3112. The dashed horizontal line indicates the WMAP measurement of the cosmic baryon budget $\Omega_{b}/\Omega_{m} = 0.167$ \citep{komatsu2011}.}
\label{fig:fgas}

\end{figure}

\begin{table*}
\caption{\it Gas Mass and Total Mass Measurements at Overdensity Radii $R_{2500}$ and $R_{500}$ } 
\begin{center}
\begin{tabular}{lcccccc}
\hline \hline 
$\Delta$ 	& $r_{\Delta}$ 	&$r_{\Delta}$ &$M_{gas}$ 	& $M_{tot}$ 	& $f_{gas}$ & Reference	\\
		& (arcsec)		& (kpc)		&($10^{13}M_{\odot}$)	& ($10^{14}M_{\odot}$)	&		&	\\
\hline\\	
2500	&	331.0$^{+16.1}_{-14.2}$&452.5$^{+26.4}_{-49.9}$ &1.68$^{+0.12}_{-0.11}$	& 1.53$^{+0.23}_{-0.19}$	& 0.109$^{+0.013}_{-0.012}$		& This work\\
2500	&	&459 &	 $-$		& $1.7\pm0.30$			& $-$	& \citet{nulsen2010}\\
\\
\hline
\\
500 & 708.8$^{+43.7}_{-36.5}$		&968.9$^{+59.7}_{-68.2}$& 4.49$^{+0.43}_{-0.45}$		& 3.00$^{+0.59}_{-0.44}$  & 0.149$^{+0.036}_{-0.032}$& This work\\
500 && 1025 	& $-$		& $2.8^{+2.4}_{-0.9}$	& $-$& \citet{nulsen2010}\\		
\\
\hline
\label{table:masses}
\end{tabular} 
\end{center}
\end{table*}

We measure gas mass, total mass and gas mass fraction at two physically meaningful overdensity radii, $r_{2500}$ and $r_{500}$. The overdensity radius $r_{\Delta}$ is defined as the radius within which the average matter density of the cluster is 
$\Delta$ times the critical density of the universe at the cluster redshift:

\begin{equation}
r_{\Delta} = \frac{M_{tot}(r_{\Delta})}{\frac{4\pi}{3}\Delta \rho_{crit}(z)}
\end{equation}

\noindent where $\rho_{crit}$  is the critical density of the Universe at the cluster's redshift. 
The measurements of  gas mass, total mass and gas mass fraction at two overdensity radii, $r_{2500}$ and $r_{500}$, are shown in Table \ref{table:masses}. 
We find that the total mass measurements obtained from our measurements are in agreement with the previous studies  (\citet{nulsen2010}; see Table \ref{table:masses} for comparison). We also estimate the total mass using the $M_{tot}$ $-\ T$ scaling relations derived from the \xmm observations of galaxy clusters  \citep{arnaud2005, mantz2010}. 
We find that the average temperature within 0.15 $r_{500}< r < r_{500}$ is $\sim$4.6 keV. The \citet{arnaud2005} $M_{tot}-T$ scaling relation produces a total mass of $M_{tot}=(3.24 \pm 0.12)\times 10^{14}M_{\odot}$ for this temperature, which is consistent with our result at the 1-$\sigma$ level. The \citet{mantz2010} $M_{tot}-T$ scaling relations produce a total mass of $M_{tot}=(3.43 \pm 1.82)\times 10^{14} M_{\odot}$ and are also consistent with our result at the 1-$\sigma$ level. 

Figure \ref{fig:fgas} shows the gas mass fraction profile. Our result agrees with the Wilkinson Microwave Anisotropy Probe (WMAP)  measured 
cosmic baryon fraction $\Omega_{b}/\Omega_{M}=0.167$at $r_{500}$  \citep{komatsu2011}.

\section{{\it RGS} Spectral Results}
The default 90 per cent of  the pulse-height (P.I.) distribution was used for the extraction of source spectra. The diffuse X-ray emission from the first-order and second-order {\it RGS} observations were then fit with an absorbed single-temperature collisional  equilibrium plasma model ({\it apec}) and AtomDB 2.0.1 (A. Foster et al. 2011, in
preparation). The best-fit temperature, abundance and the goodness of the fit  is given in Table \ref{table:rgsFitResults}. 
Since C-statistic does not provide a direct estimate of the goodness-of-fit and we asses the goodness of the fit using
 the $\chi^{2}$ value we calculated from the best-fit obtained from C-statistics.
The final $\chi^{2}$ values reported in Table \ref{table:rgsFitResults} were obtained from the best fit determined by C-statistics.

\begin{table*}
\caption{\it  {\it RGS}-fit Results for Two Different Region Selections}
\centering
\scriptsize 
\begin{tabular}{lccccccc}
\hline \hline 
 P.I. Width$^{*}$ 		& Source Extent	& kT	& Abundance &  Red. $\chi^{2}$ \\ 
 (per cent)			& $''$			& (keV)	&	 &(d.o.f.) 	\\	
			\hline\\
 90				& 37.9 		&  3.82 $\pm$ 0.21  & 0.78 $\pm$ 0.02 & 1.36 (3287)\\		
 50				& 30.6		& 3.52 $\pm$ 0.37  & 0.86 $\pm$ 0.17 & 1.60 (2468)\\
	
\\
\hline
\label{table:rgsFitResults}
\end{tabular} 
\\
$^{*}$The inclusion region of the pulse-height mask.
\end{table*}

\label{sec:rgsResults}

\subsection{An Upper Limit to the Source Extent}

\begin{figure}
\centering
\includegraphics[angle=-90,width=8.7cm]{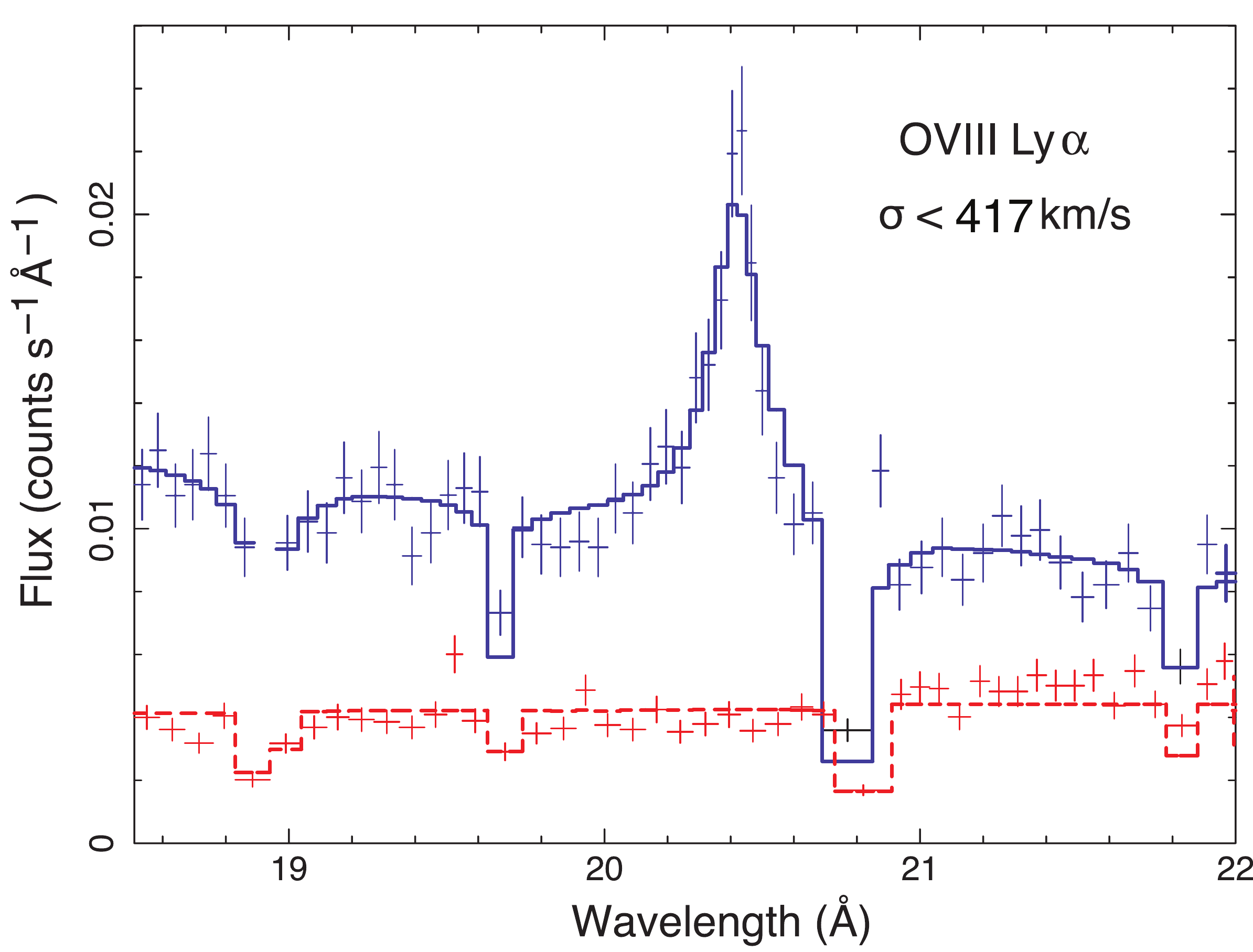}
\caption{ Close up view of oxygen VIII Ly$\alpha$ line obtained from high resolution XMM-Newton {\it RGS} observations of A3112. The best fit Gaussian line is shown in blue. The background spectrum is shown in red with the best fit model (dashed red line). This fit produces total $\chi^{2}$ 64.3 for 50 degrees of freedom for source spectrum and  89.8 for 43 degrees of freedom for background spectrum.
}
\label{fig:rgs_Lalpha}
\end{figure}

The two different region selection filtering was used for the extraction of source spectra (the default 90 per cent and 95 per cent). The best-fit temperature, abundance and the goodness of the fit for these two different aperture are given in Table \ref{table:rgsFitResults}. 
In order to determine the source extent observed by the {\it RGS} we determined the line width of the O VIII Ly$\alpha$ line without applying the spectral broadening convolution model to the response files for two different pulse height extraction mask. The best fit  model to data for O VIII Ly$\alpha$ observed line energy (0.607 keV, 20.44\AA) and width (2.26$\times10^{-3}$ keV, 0.043\AA) is shown in Figure \ref{fig:rgs_Lalpha}. The  observed line emission has been broadened by a combination of the thermal broadening of the ICM, turbulent motions within the observed region and  the spatial extent of the source along the dispersion direction. Assuming that the O VIII line is intrinsically narrow and only broadened by the spatial extent of the source we placed an upper limit to the source extent observed by {\it RGS} to be $\sim38''$ ($\sim$ 52 kpc) using Equation \ref{eqn:broadEqn}. This result shows that the {\it RGS} is only sensitive to emission from the cluster 
core, and retains useful resolution in these observations.
The temperature and metal abundances measured by the {\it RGS} spectra are consistent with the central bin temperature obtained from {\it {\it MOS}} and {\it PN} observations (see Figure \ref{fig:epicTAP}).

\subsection{Elemental Abundances in the Core}

We fit the {\it RGS} first order and second order spectra with a single temperature thermal model (\textit{vapec}).
The O, Ne, Mg, Si and Fe abundances were allowed to vary
while all other metal  abundances were tied to Fe abundance. 
The normalizations  were allowed to vary between first second and second order spectra.
The best fit temperature and elemental abundances obtained from this fit are shown in Table \ref{table:individualAbund} with the goodness of the fit.

\begin{table*}
\caption{\it Temperature and Abundance Best-fit Parameters of {\it RGS} Data Extracted From the 90 per cent PSF aperture.} 
\centering
\footnotesize 
\begin{tabular}{lcccc}
\hline \hline 
Parameter		&vapec	& 2 $\times$ vapec  &vapec + vmcflow& 6 $\times$ vapec\\
			&		&				& $kT_{min}$ free & \\
\hline
\\
kT (keV)				& 3.89 $\pm$ 0.26 	& 4.75 $\pm$ 0.25		& 4.78 $\pm$ 0.33 &4.0, 2.0, 1.0, 0.5,\\
kT$_{min}$ (keV)				&	$-$			& 1.01 $\pm$ 0.03 		& 0.52 $\pm$0.07 & 0.25, 0.0808\\
O ($A_{\odot}$)			& 0.74 $\pm$ 0.14 	& 0.90 $\pm$ 0.13		& 0.72 $\pm$ 0.12 & 0.75 $\pm$ 0.03\\
Ne ($A_{\odot}$) 		& 2.10 $\pm$ 0.40 	& 2.52 $\pm$ 0.35		& 2.36 $\pm$ 0.30& 2.21 $\pm$ 0.21\\
Mg ($A_{\odot}$)		& 1.10 $\pm$ 0.26	& 1.45 $\pm$ 0.30		&1.46 $\pm$ 0.37&1.70 $\pm$ 0.20\\
Fe ($A_{\odot}$)		& 0.90 $\pm$ 0.12	& 1.33 $\pm$ 0.11		&1.57 $\pm$ 0.30 &1.28 $\pm$ 0.024\\
Mass deposition rate	& 	$-$			&  	$-$				& 43.5 $\pm$ 3.5 & $-$\\
Norm ($\times\ 10^{-4}$)	 & 127.7 $\pm$	0.17	& 119.5 $\pm$ 2.3		& 94.5 $\pm$ 6.4 & See Figure \ref{fig:rgsEmissionMeasure}\\
					& $-$				& 2.76 $\pm$ 0.55 & $-$ & $-$\\
$\chi^{2}$ (d.o.f.) of source spectra & 4149.5 (2732)	& 4067.0 (2730)	& 4122.0 (2731)	& 4039.6 (2727)	\\
$\chi^{2}$ (d.o.f.) of background spectra 	& 2167.1 (1332)& 2382.4 (1332)& 2410.9 (1332)	& 2370.9 (1332) \\
\\
\hline
\label{table:individualAbund}
\vspace{5mm}
\end{tabular} 
\end{table*}

We find that the temperature of the gas in the cluster core obtained from {\it RGS} spectra agrees with the temperature of the central bin (0 $-$ 30$''$) of EPIC observations.
The abundance measurements of Fe, O and Si in the innermost 38$''$ region obtained from {\it RGS} observations are consistent with the abundance in the central bin of EPIC observations at 1-$\sigma$ level (see Figure \ref{fig:epicFeOSi}).

\subsection{Multi-temperature Fit}

In order to test if there is a multi-component  gas in the core as found in other clusters, e.g., Centaurus \citep{takahashi2009,sanders2008}, we fit {\it RGS} spectra with two \textit{vapec} models each with a variable temperature, normalization. The Ne, O, Mg and Fe metallicities were allowed to vary independently while  
the Si, Ca, Al, S , Ar and Ni metallicites were linked to Fe and tied between two different {\it vapec} components.

\begin{figure}[ht!]
\centering
\includegraphics[angle=-90,width=9.4cm]{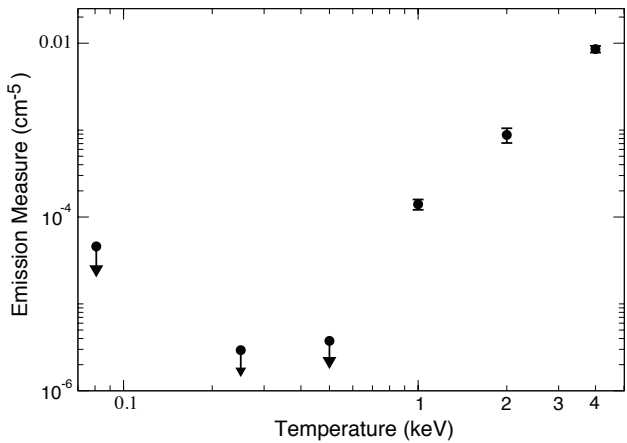}
\caption{Emission measure as a function of temperature obtained from {\it RGS} observations. {\it RGS} spectra was fit with six \textit{vapec} models with fixed temperatures of 4.0, 2.0, 1.0, 0.5, 0.25, and 0.0808 keV. The best fit abundances and goodness of the fit are shown in Table \ref{table:individualAbund}. }
\label{fig:rgsEmissionMeasure}
\end{figure}

We also extended the two-temperature model to six temperature components with fixed temperatures ranges 4.0, 2.0, 1.0, 0.5, 0.25 and 0.0808 keV to fit the {\it RGS} spectra. We coupled all metal abundances for each model to reduce the number of free parameters but allowed each normalization to vary.
Table \ref{table:individualAbund} shows the best-fitting parameters for these six \textit{vapec} models. The  temperature weighted emission measure for each component is plotted as a function of temperature in Figure  \ref{fig:rgsEmissionMeasure}. Figure \ref{fig:rgsEmissionMeasure} indicates 
the apparent lack of any gas below 1 keV in the {\it RGS} results. A3112 is particularly different from the previous results reported for Centaurus, HGC 62, A262 and A3581 \citep{sanders2008, sanders2010b} since {\it RGS} observations of these clusters show a multiphase gas at lower temperatures ($\sim$ 0.5 keV).

\subsection{Fitting the Cooling Flow}

We also determine the cooling rate of the gas as a function of temperature using the latest AtomDB v2.0.1. The XSPEC model \textit{vmcflow} models gas cooling between two temperatures and gives the normalization as a mass deposition rate. Here  the mass deposition rate  is an upper limit to the gas which cools below X-ray temperatures determined from the X-ray spectra. We constructed a  model composed of a vapec model and a cooling flow model components (\textit{vmcflow}) to fit the {\it RGS} spectra. We left the $kT_{min}$ parameter free in order to determine the minimum temperature that gas is cooling to whereas the maximum temperature was linked to the temperature of the \textit{vapec} model.
The abundances of each element was allowed to vary while the redshift was fixed to the value reported in Table \ref{table:individualAbund}. The mass deposition rate we obtained from this fit is  43.5 $\pm$ 3.5 M$_{\odot}$ yr$^{-1}$ .
We also found that there is no evidence for the  emission below a cut-off at  $\sim$1 keV. The lack  of emission from cool gas indicates that this emission property is general for clusters with a cD galaxy.

From Einstein \citep{edge1992} and ROSAT \citep{allen1997}  imaging data with little spectral information a large total mass deposition rate has been reported for A3112 of 415 M$_{\odot}$ yr$^{-1}$. 
\citet{odea2008} reported that the total mass deposition rate was much less than previously claimed with $10^{+7}_{-5}$ M$_{\odot}$ yr$^{-1}$ being  consistent with the {\it Chandra} and {\it XMM-Newton} X-ray spectra. 
The level of cooling we report in this work is also consistent with the star formation rate of 4.2 $M_{\odot} \ yr^{-1}$ measured from the Spitzer IR data \citep{odea2008}.

\subsection{An Upper Limit on Turbulence }

The turbulent gas motion in the ICM can  be generated  by frequent  mergers \citep{ascasibar2006}, the motion of galaxies \citep{conroy2008}, SN explosions and, AGN outbursts \citep{mcnamara2007}. The line widths of heavy elements are dominated by thermal broadening. Very low  ($\sim$50$-$150 km s$^{-1}$)  turbulent motions  can  affect the magnetic field orientation in the ICM and increase the effective level of thermal conduction \citep{ruszkowski2010}. The dissipation of turbulent gas motions has been proposed as one of the mechanisms that heats the ICM in the core of clusters and prevents the gas from extreme cooling \citep{rebusco2008, ruszkowski2010}.

The earliest direct upper limit on turbulence in the X-ray waveband was placed by \citet{sanders2010a} examining the \xmm {\it RGS} spectra  of  A1835. The limit on the total broadening, including the turbulent component, of 274 km s$^{-1}$ was reported at the 90 per cent level. It was also reported that the ratio of turbulent to thermal energy density in the core of A1835 is less than 13 per cent. 
\citet{sanders2010a,sanders2011} placed upper limits on the velocity broadening examining the \xmm {\it RGS} observations of 62 galaxy clusters, galaxy groups, and elliptical galaxies. After the addition of  a 50 km s$^{-1}$ systematic uncertainty it was reported that five of these sources to have less than $\sim$20 per cent of the
thermal energy density is related to turbulence, with velocity upper limits $\leq$220 km s$^{-1}$.

\begin{table}
\caption{\it Measured Value of the total velocity broadening. If line smearing effect due to the spatial extended of the source is ignored this velocity corresponds to turbulent velocity.}
\centering
\footnotesize 
\begin{tabular}{lcccc}
\hline \hline 
Line			& $v_{\|} $	& Energy Fraction\\
			& (km s$^{-1}$)	& (\%)	\\
\hline
\\
OVIII line$-$only	&804$^{+353}_{-110}$ 	& 52$^{+17}_{-2}$ 	\\
	\\
{\it bvapec} (all lines) & 834$^{+119}_{-105}$&  54$\pm$2 	\\
\\
\hline
\label{table:velocityBroadening}
\end{tabular} 
\end{table}

We examine the deep {\it RGS} observations of A3112 and  can place an upper limit  on velocity broadening of the heavy metals  in the ICM.
The total line width due to thermal motions of ions and turbulence is, 

\begin{equation}
\sigma^{2}= \sigma_{turb}^{2}+\sigma_{therm}^{2}.
\end{equation}

\noindent the line of sight component of line width due to turbulence is 

\begin{equation}
\sigma_{turb} = \nu_{0}\frac{v_{turb,\|}}{c}
\end{equation}

\noindent and of the line of sight component of line width due to thermal motion of ion \textit{i} is 

\begin{equation}
\sigma_{therm}^{2}=\frac{kT_{e}}{m_{i}} \left( \frac{\nu_{0}}{c}\right)^{2}.
\end{equation}

\noindent Here $\nu_{0}$ is the rest frequency of the line of interest, $T_{e}$ is the electron temperature, $m_{i}$ is the mass of the ion \textit{i}, and $c$ the speed of light \citep{rebusco2008}. Then the line of sight velocity of turbulence is 

\begin{equation}
v_{turb,\|} = c \left( \frac{\sigma^{2}}{\nu_{0}^{2}} - \frac{kT_{e}}{m_{i}c^{2}}\right)^{1/2}.
\label{eqn:turbVel}
\end{equation}

The energy density contribution due to turbulence can be calculated via,

\begin{equation}
\frac{\epsilon_{turb}}{\epsilon_{therm}}=\mu m_{p} \frac{v_{turb,\|}^{2}}{kT_{e}}
\label{eqn: tubulenceEnerg}
\end{equation}

\noindent where $\mu$ is the mean molecular weight, $v_{turb,\|}$ is the line of sight velocity, 
and $m_{p}$ is the proton mass \citep{werner2009}. 

\begin{table}
\caption{\it 90 Per cent Upper Limits to Turbulent Velocity Broadening ($v_{turb,\|}$). \textit{rgsrmfsmooth} was used for removing the effect of the line broadening due to the extended nature of the source.}
\centering
\footnotesize 
\begin{tabular}{lcccc}
\hline \hline 
Line			& $v_{turb,\|}$	& Energy Fraction\\
			& 	(km s$^{-1}$)	& (\%)	\\
\hline
\\
OVIII line$-$only		& 417	& 24\\
\\
{\it bvapec} (all lines) 	& 206  	& 6 \\
\\
\hline
\label{table:turbvelocityBroadening}
\end{tabular} 
\end{table}

In order to test an extreme case we first assume that the line broadening due to source extent is negligible and lines are broadened purely by the thermal motions of ions. The best fit total broadening obtained from the RGS spectra was used in Equation \ref{eqn:turbVel} to calculate the velocity of turbulence. The energy contribution of turbulence to the total energy was calculated using Equation \ref{eqn: tubulenceEnerg}. 
We found that the velocity of broadening is 804$^{+353}_{-110}$ km s$^{-1}$ with an energy contribution of 52$^{+17}_{-2}$ per cent by examining only OVIII line region between 18.5 \AA~and 22 \AA~wavelength range. We also fit {\it RGS} first order and  second order spectra with \textit{bvapec} model
which utilizes the AtomdDB  2.0.1 atomic data base and includes a velocity fit parameter that accounts for line broadening \citep{sanders2010a}. The entire wavelength band (7 $-$ 28\AA) we used for this analysis gives a measurement  of 834$^{+119}_{-105}$ km s$^{-1}$ (see Table \ref{table:velocityBroadening}).  

We then correct for the line broadening due to the source extent as described in \S\ref{sec:RGSProcessing} by convolving response matrices  with the {\it MOS} image.  
This fit produces a total $\chi^{2}$ of 4061.4 for 2650 dof for the source spectra and  2866.1 for 1537 dof for the background spectra.
We find a 90 per cent upper limit to turbulence velocity of 206 km s$^{-1}$ with an energy density contribution of 6 per cent for the entire {\it RGS} wavelength band (7 $-$ 28\AA). Using only the OVIII line (18.5 \AA~and 22 \AA) wavelength range we obtain a 90 per cent upper limit of 417  km s$^{-1}$ with a energy contribution of less than 24 per cent (see Table \ref{table:turbvelocityBroadening} and Figure \ref{fig:rgs_Lalpha}). Total $\chi^{2}$ we obtained from this fit is 64.3 for 50 dof for the source spectrum and  89.8 for 43 dof for the background spectrum.

\begin{table*}
\caption{\it 90 per cent Upper Limits to Line Fluxes and Metal Abundance obtained from {\it RGS} Spectra}
\centering
\footnotesize 
\begin{tabular}{lcccc}
\hline \hline 
Line		& Transition	& $\lambda_{rest}$		& Flux		& Abundance\\
		&	& ($\AA$)			& 	( $10^{-6}$ photons $cm^{-2} s^{-1}$)& $<$	\\
\hline
Fe XVII 	& 3d $\rightarrow$ 2p	& 15.014			& 1.00	& 4.98$\times10^{-4}$\\
Fe XVII	& 3s $\rightarrow$ 2p	& 16.780			& 0.84	& 6.89$\times10^{-4}$\\
Fe XVII	& 3s  $\rightarrow$ 2p	& 17.096			& 6.97	& 4.25$\times10^{-3}$\\
Fe XVIII	& 3d  $\rightarrow$ 2p	& 14.208			& 5.45	& 5.58$\times10^{-3}$\\
Fe XVIII	& 3s  $\rightarrow$ 2p	& 16.071			& 2.40	& 4.39$\times10^{-3}$\\
\hline
\label{table:lineFluxes}
\vspace{5mm}
\end{tabular} 
\end{table*}

 We repeat the analysis of modifying the response matrices convolving with a \textit{Chandra} image obtained from the offset {\it ACIS-I} observations to correct for the line smearing due to the spatial structure of the source. We obtain a total $\chi^{2}$ of 4003.1 for 2650 dof from the source spectra from this fit.
Background spectra produce a total $\chi^{2}$ of 2840.0 for 1537 dof.
This analysis produces a 90 per cent upper limit to the velocity broadening of 225 km s$^{-1}$ for the entire {\it RGS} wavelength band (7 $-$ 28\AA). The vignetting and off-axis effects of {\it Chandra ACIS-I } image could partially be responsible for the difference between upper limits to the velocity broadening. For self-consistency in the analysis we report upper limits obtained using the {\it MOS} image as described in \S \ref{sec:xmmEpicDataAnalysis}. We also note that {\it rgsrmfsmooth} routine which is used to correct for broadening due to the spatial structure of the source makes an approximation that the source spectrum does not change as a function of spatial position. 

Non-thermal pressure support from bulk motions in clusters cores plays an important role in the thermal properties of galaxy clusters. In particular, the Sunyaev Zel'dovich (SZ) effect  observables for cluster surveys, such as the Atacama Cosmology Telescope, the South Pole Telescope and Planck, CARMA/SZA are affected by these processes \citep{nagai2007,shaw2010, battaglia2011}. 
Several earlier simulations showed that the kinetic pressure from bulk motions contributes a small ( $\sim$ 5\% $-$15\%) but still significant amount of energy within cluster core scales in relaxed clusters, which are consistent with out upper limit \citep{dolag2005,lau2009}. More recent works, performing high-resolution simulations with the AMR code ENZO simulations of galaxy clusters have shown that the energy contribution of turbulent motions, $E_{turb}/E_{therm}$, is $\leq$5 per cent in the compact cores of relaxed clusters, are consistent with our upper limit \citep{iapichino2008, lau2009, burns2010,vazza2011}.

\subsection{Absence of Low Temperature Component and An Upper Limit on Metal Abundance}

The soft excess was found earlier in A3112  in the XMM-Newton, Chandra and Suzaku data by  \citet{nevalainen2003}, \citet{bonamente2007}, and \citet{lehto2010}.
Interpretation of the soft excess emission has been under debate. One possible explanation is that this excess emission originates from a thermal warm gas  associated with the cluster (kT $\sim$ 0.1$-$1 keV)  \citep{lieu1996, bonamente2007}, which indicates the presence of an additional phase in the intergalactic medium. \citet{bonamente2007} have placed an upper limit on metal abundance of $\leq$ 0.01 by fitting  the soft excess emission detected from the XMM-MOS observations with an additional single temperature thermal model.
\begin{figure}[h!]
\centering
\includegraphics[angle=-0,width=8.5cm]{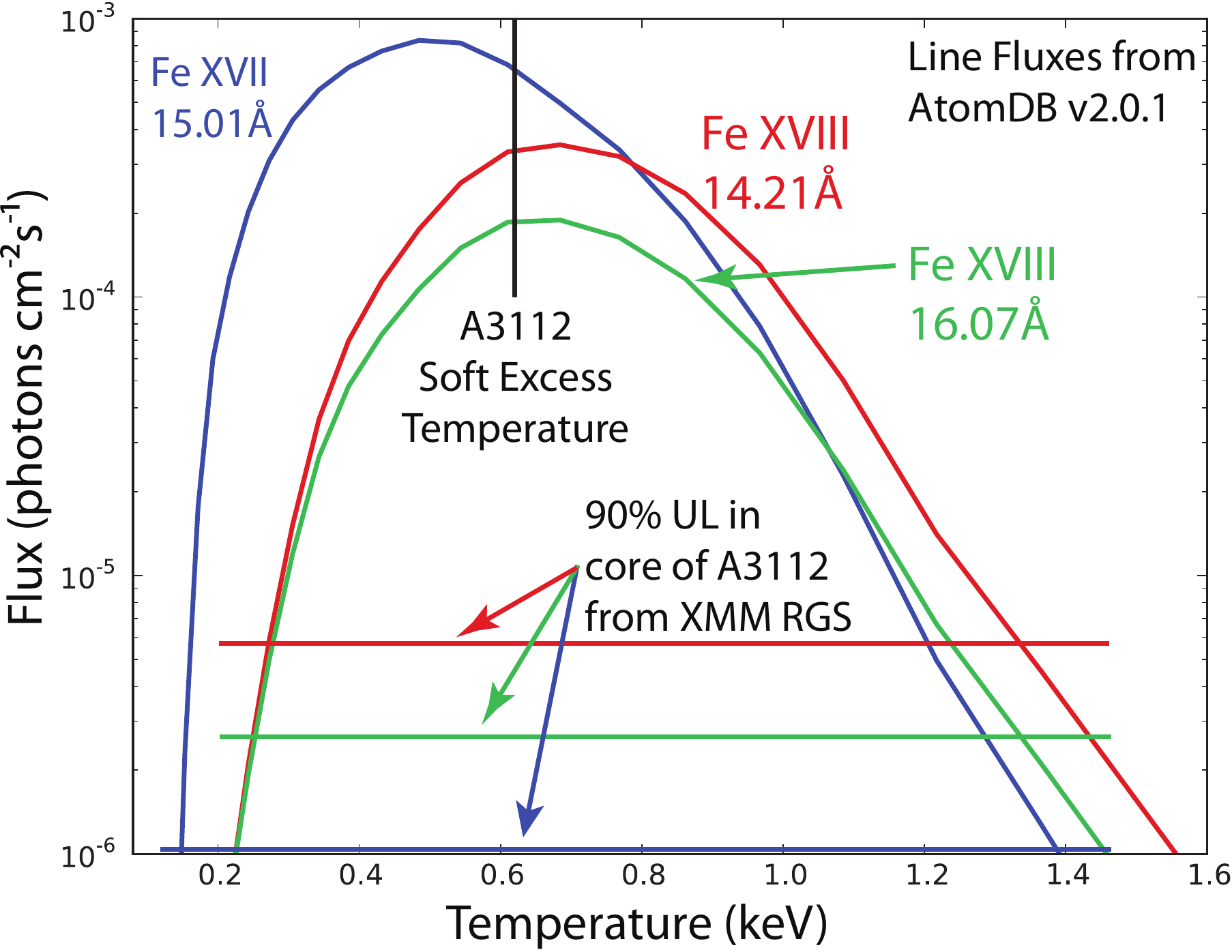}
\caption{ 90 per cent upper Limits to the Fe XVII and Fe XVIII Line Fluxes obtained from the {\it RGS} observations A3112 are shown in green, red and blue straight lines. The curves show the expected  line fluxes for \citet{bonamente2007}  temperature  of 0.62 keV plasma, shown in vertical black line, with normalization of 1.05$\times 10^{-2}$ obtained from AtomDB 2.0.1 database (A. Foster et al. 2011, in preparation).
}
\label{fig:rgs_softExcess}
\vspace{3mm}
\end{figure}

We, therefore, searched for the signatures of $< 1$\,keV gas in the {\it RGS} spectra in order to test the hypothesis of the thermal origin of the soft excess detected from this cluster and to determine a new upper limit on the metal abundance by assuming that the soft excess follows the cluster's surface brightness. Since the Fe XVII and Fe XVIII lines are sensitive indicators of a plasma with a temperature of $<$ 1 keV, we determined the Fe XVII  line flux limits, are shown in Table \ref{table:lineFluxes}. To put a limit on any emission from Fe L shell lines, we added a $\delta$ function to the model at the energy where the Fe XVII and Fe XVIII lines would be expected. The 90 per cent upper limits to the Fe XVII and Fe XVIII fluxes are shown in Table \ref{table:lineFluxes}. These lines are emitted by cooler gas (e.g. Fe XVII, XVIII) come from more compact regions, and therefore the limits we report are for compact emission from Fe XVII, XVIII lines. The limits on Fe XVII, XVIII lines from extended emission are presumably larger.

These upper limits were then compared with the line fluxes estimated using AtomDB
v2.0.1. The fluxes were estimated assuming an optically-thin
low-density plasma. 
Best-fit values for the {\it MOS} observations from \citet{bonamente2007} were used to convert from emissivities 
to line fluxes: the geometric normalization factor of 1.05$\times 10^{-2}$ and column density of $2.6\times10^{20}$ cm$^{-2}$.
The emissivity of the Fe XVII and Fe XVIII ions peaks in the 0.5$-$0.7 keV
temperature range, with predicted fluxes 1 to 3 orders of magnitude
higher than the observed 90 per cent upper limits obtained from {\it RGS} spectra at the plasma temperature 0.62 keV given by \citet{bonamente2007}, as shown in Figure \ref{fig:rgs_Lalpha}. The upper limits determined from Fe XVII and Fe XVIII line fluxes are shown in Table \ref{table:lineFluxes}. 
This implies that the product of the abundance and normalization (A$\times$N) must be 100 times
less. These abundance and normalization products we obtained from {\it RGS} spectra allow us to place a new upper limit on the metal abundance. These upper limits we obtained from {\it RGS} spectra indicate that 0.62 keV plasma associated with the cluster is either very tenuous, to the point of non-existence, or has a very low metallicity less than $4.98\times10^{-4}$ in the central $<38''$ region of the cluster.

 \section{Conclusion}
\label{sec:conclusion}
In this work we studied the X-ray emission of a nearby rich cluster A3112 from deep XMM-Newton EPIC and {\it RGS} observations. The XMM-Newton EPIC data processing and background modeling were carried out with the most up-to-date methods \citep{kuntz2008,snowden2008}. The temperature and abundance profiles obtained from our combined {\it MOS} and {\it PN} fits are consistent with those measured  by \citet{snowden2008}.  We find that {\it MOS} temperature measurements are systematically higher than {\it MOS}+{\it PN} combined temperatures. This systematic discrepancy in the wide band temperatures of up to 7 per cent was also observed  by \citet{nevalainen2010}. The abundance measurements obtained from the {\it MOS} observations are in good  agreement with those obtained from the {\it MOS}+{\it PN} observations and the \citet{snowden2008} results. From the EPIC fits we find that the iron abundance gradient shows a significant increase towards the center of the cluster while the O abundance profile is also centrally peaked but has a shallower distribution than that of Fe. 
The centrally peaked trend of Fe is qualitatively consistent with the idea that the dominant contribution from SNe Ia is towards the cluster center, while the distribution of SNe II remains more or less uniform \citep{kaastra2001,tamura2004}.  This result may also indicate that iron is still being added to the ICM specifically in the core by the supernovae SNe Ia.

We use the \citet{bulbul2010} model to examine the cluster properties as a function of radius. The projected density and temperature are obtained  from the {\it MOS} spectra. We then deproject the density profile obtained from XSPEC normalizations and project the three dimensional temperature model to fit the {\it MOS} data. 
We demonstrated that the \citet{bulbul2010} model accurately describes the number density and temperature profiles from the obtained from the \xmm X-ray data. At $r_{500}$, the total mass $M_{500}=(3.00^{+0.59}_{-0.44})\times 10^{14} M_{\odot}$, is in agreement with the results of \citet{nulsen2010}. The total mass also obeys the empirical $M - T$ scaling relations found in general massive galaxy clusters \citet{arnaud2005,mantz2010}. The gas mass fraction is $f_{gas}= 0.149^{+0.036}_{-0.032}$ at $r_{500}$, and is consistent with the seven-year WMAP results \citep{komatsu2011}.

We also searched for the signatures of $< 1$\,keV gas. The comparison of flux limits on the Fe XVII and Fe XVIII  lines in the {\it RGS} spectra with the AtomDB line fluxes is obtained using the \citet{bonamente2007} values for the plasma temperature, column density,
abundance, and the geometric normalization factor. The upper limits on the Fe XVII and Fe XVIII  lines we obtained from {\it RGS} observations allow us to place an upper limits on metal abundance in the central 38$''$ region of A3112. These upper limits indicate that either there is no spectral evidence for cooler gas associated with the cluster with temperature below 1.0 keV in the central 38$''$ ($<$ 52 kpc) central region of A3112 or either it has a very low metallicity ($< 4.98\times10^{-4}$).

We also determined the emission measure distribution as a function of temperature, detecting or limiting the amount of cool gas in the cluster core. Alternatively, the emission should have a cut-off at  $\sim$1 keV.  The lack  of emission from cool gas has been found also in similar systems, such as A1835 \citep{peterson2001}, Sersic 159-03 \citep{kaastra2001}, and A1795 \citep{tamura2001}, indicating that this emission property is general for clusters with a cD galaxy.

We also placed an upper limits to turbulent gas motions directly using high-resolution {\it RGS} X-ray spectra of the compact core of A3112.
The upper limit of 206 km s$^{-1}$ to turbulent motions with less that 6 per cent  contribution to the total energy that we found in the core of the relaxed cluster A3112 is compatible with the values of $\sim$6 per cent of pressure in gas motions, calculated using high-resolution Eulerian simulations \citep{lau2009, burns2010,vazza2011}. These upper limits were obtained based on an assumption that the source spectrum does not change as a function of spatial position. The low limit we measured on turbulent motions from the core of a relaxed galaxy cluster A3112 is encouraging for the  cosmological studies using SZ cluster surveys. In the near future, the launch of Astro-H will provide an X-ray calorimeter with high energy resolution of $\sim$4.5 eV. This detector will produce measurements of lines widths with high accuracy revolutionizing the measurement of turbulent velocities in the ICM.

\section*{Acknowledgments} The authors thank 
Steve Snowden and Helen Russell  for kindly providing help on the {\it ESAS} software and deprojection methods. We also thank the anonymous referee, Maxim Markevitch, Paul Nulsen, Max Bonamente, and Jelle de Plaa for providing useful suggestions and comments on the manuscript. We gratefully acknowledge support for this research from NASA XMM-Newton grant NNX09AP92G and NASA ROSES-ADP grant NNX09AC71G.

\end{document}